\documentclass[reprint, amsmath, amssymb, aps, nobibnotes, nofootinbib, longbibliography]{revtex4-2}

\usepackage[utf8]{inputenc}
\usepackage[english]{babel}

\usepackage{graphicx}
\usepackage{hyperref}
\usepackage{bm}
\usepackage{amsmath}
\usepackage{amsfonts}
\usepackage{amssymb}
\usepackage{amsthm}
\usepackage{mathrsfs}
\usepackage[mathcal]{euscript}
\usepackage{ulem}

\usepackage{epsfig}
\usepackage{float}
\usepackage{latexsym}
\usepackage{psfrag}
\usepackage{nicefrac}%inline fractions
\newcommand{\be}{\begin{equation}}
\newcommand{\ee}{\end{equation}}
\newcommand{\bea}{\begin{eqnarray}}
\newcommand{\eea}{\end{eqnarray}}

\newcommand{\RNumb}[1]{\textbf{\uppercase\expandafter{\romannumeral #1\relax}}}
\newcommand{\nn}{\nonumber}
\newcommand{\eps}{\varepsilon}
\newcommand{\up}{\text{up}}
\newcommand{\down}{\text{down}}
\newcommand{\MS}{\text{MS}}
\newcommand{\AS}{\text{AS}}
\newcommand{\m}{\text{m}}
\newcommand{\gen}{\text{gen}}

\newcommand{\extgen}{^\text{ext}_\text{gen}}
\newcommand{\GN}{G_{\mathrm{N}}}
\newcommand{\BH}{\text{B-H}}
\newcommand{\tbreak}{t_\text{break}}
\newcommand{\LateTimes}{\Big\rvert_{\begin{subarray}{l}\text{late}\\\text{times}\end{subarray}}}
\newcommand{\InTextLateTimes}{\rvert_{\begin{subarray}{l}\text{late}\\\text{times}\end{subarray}}}
\newcommand{\InterTimes}{\Big\rvert_{\begin{subarray}{l}\text{inter.}\\\text{times}\end{subarray}}}
\newcommand{\EarlyTimes}{\Big\rvert_{\begin{subarray}{l}\text{early}\\\text{times}\end{subarray}}}
\newcommand{\ba}{\mathbf{a}}
\newcommand{\bb}{\mathbf{b}}
\newcommand{\bq}{\mathbf{q}}
\newcommand{\bp}{\mathbf{p}}
\newcommand{\bw}{\mathbf{w}}
\newcommand{\bx}{\mathbf{x}}
\newcommand{\by}{\mathbf{y}}

\newcommand{\skipline}{\vspace{\baselineskip}}

\usepackage{soul}
\usepackage{comment}
\usepackage{xcolor}

\definecolor{darkblue}{rgb}{0,0,1}
\newcommand{\IA}{\textcolor{darkblue}}

\definecolor{dgreen}{rgb}{0,0.6,0}

\definecolor{darkraspberry}{rgb}{0.9,0.,0.3}

\newcommand{\AB}{\textcolor{cyan}}

\definecolor{aquamarine}{rgb}{0.8,0.0,0.8}
\newcommand{\TR}{\textcolor{aquamarine}}

\definecolor{ddgreen}{rgb}{0,0.8,0}

\begin{document}

\title{Infrared Regularization and Finite Size Dynamics of Entanglement Entropy in~Schwarzschild Black Hole}

\author{D.S.~Ageev$^{a}$}
\email{ageev@mi-ras.ru}

\author{I.Ya.~Aref'eva$^{a}$}
\email{arefeva@mi-ras.ru}

\author{A.I.~Belokon$^{a}$}
\email{belokon@mi-ras.ru}

\author{A.V.~Ermakov$^{b}$}
\email{alexey.ermakov@manchester.ac.uk}

\author{V.V.~Pushkarev$^{a}$}
\email{pushkarev@mi-ras.ru}

\author{T.A.~Rusalev$^{a}$}
\email{rusalev@mi-ras.ru}

\affiliation{${\hspace{0pt}}^{a}$Steklov Mathematical Institute, Russian Academy of Sciences,\\ Gubkin str. 8, 119991 Moscow, Russian Federation \\ ${\hspace{0pt}}^{b}$Department of Physics and Astronomy, University of Manchester, \\ Oxford Road, M13 9PL Manchester, England, United Kingdom}

\date{\today}

\begin{abstract}
In this paper, infrared regularization of semi-infinite entangling regions and island formation for regions of finite size in the eternal Schwarzschild black hole are considered. We analyze whether the complementarity property and pure state condition of entanglement entropy can be preserved in the given approximation. We propose a special regularization that satisfies these two properties. With regard to entangling regions of finite size, we derive two fundamental types of them, which we call ``mirror-symmetric''~(MS) and ``asymmetric''~(AS). For MS regions, we discover a discontinuous evolution of the entanglement entropy of Hawking radiation due to finite lifetime of the island. The entanglement entropy of matter for semi-infinite regions in two-sided Schwarzschild black hole does not follow the Page curve. The lifetime of AS regions is bounded from above due to the phenomenon that we call ``Cauchy surface breaking''. Shortly before this breaking, the island configuration becomes non-symmetric.  For both types of finite regions, there is a critical size, below which the island never dominates. For regions smaller than some other critical size, the island does not emerge. Finally, we show that the island prescription does not help to solve the information paradox for certain finite regions.
\end{abstract}

\maketitle

%\tableofcontents
%%%%%%%%%%%%%%%%%%%%%%%%%%%%%%%%%%%%%%%%%%%%%%%%

\section{Introduction}

Hawking radiation is a phenomenon which opens up the window into the world of quantum effects emerging in gravity~\cite{Hawking1975particle, Hawking:1976ra}. Years ago, Page showed~\cite{Page:1993wv, Page:2013dx} that a detailed comparison of the thermodynamic entropy of black holes and the entanglement entropy of their radiation results in the observation that the latter exhibits an unlimited growth and finally exceeds the Bekenstein-Hawking entropy. This is in contradiction with the expected time dependence of the entanglement entropy visualized by the Page curve, which should start decreasing after a certain point, called the Page time. The descending part of the Page curve is difficult to interpret in a straightforward way, and recently an ``island proposal'' was introduced to explain the stoppage of the entanglement entropy growth~\cite{Penington:2019npb, Almheiri:2019psf, Almheiri:2019hni}. This proposal is applicable to systems with dynamical gravitational degrees of freedom. The conjecture of modifying entanglement entropy in the presence of dynamical gravity has attracted a lot of attention in recent years~\cite{Davies:1976hi, Good:2016atu, Chen:2017lum, Good:2019tnf, Almheiri:2019yqk, Rozali:2019day, Penington:2019kki, Almheiri:2019qdq, Gautason:2020tmk, Anegawa:2020ezn, HIM, Sully:2020pza, Hartman:2020swn, Krishnan:2020oun, Alishahiha:2020qza, Geng:2020qvw, Chen:2020uac, Almheiri:2020cfm, Dong:2020uxp, Balasubramanian:2020xqf, Sybesma:2020fxg, Chen:2020hmv, Ling:2020laa, Matsuo:2020ypv, Goto:2020wnk, Akal:2020twv, Geng:2020fxl, Karananas:2020fwx, Wang:2021woy, Kawabata:2021hac, Reyes:2021npy, Geng:2021wcq, Bhattacharya:2021jrn, Kim:2021gzd, Aalsma:2021bit, Geng:2021iyq, Lu:2021gmv, Akal:2021foz, Yu:2021cgi, Geng:2021hlu, Ahn:2021chg, Ageev:2021ipd, Balasubramanian:2021xcm, Cao:2021ujs, Fernandez-Silvestre:2021ghq, Azarnia:2021uch, Bhattacharya:2021dnd, Arefeva:2021kfx, He:2021mst, Ageev:2019xii, Omidi:2021opl, Bhattacharya:2021nqj, Geng:2021mic, Yu:2021rfg, Suzuki:2022xwv, Hu:2022ymx, Hu:2022zgy, Arefeva:2022cam, Gan:2022jay, BasakKumar:2022stg, Azarnia:2022kmp, Akal:2022qei, Afrasiar:2022ebi, Anand:2022mla, Ageev:2022hqc, Djordjevic:2022qdk, Goswami:2022ylc, RoyChowdhury:2022awr, Geng:2022dua}. Entanglement islands have been studied in the setups of two-dimensional gravity~\cite{Penington:2019npb, Almheiri:2019qdq, Penington:2019kki, Almheiri:2019yqk, Almheiri:2020cfm, Chen:2020uac, Chen:2020hmv, Karananas:2020fwx, Arefeva:2021kfx}, boundary CFT~\cite{Rozali:2019day, Sully:2020pza, Geng:2021iyq, Ageev:2021ipd, Geng:2021mic, Suzuki:2022xwv, Hu:2022ymx, Hu:2022zgy, BasakKumar:2022stg, Afrasiar:2022ebi, Geng:2022dua} and moving mirror models~\cite{Davies:1976hi, Good:2016atu, Chen:2017lum, Good:2019tnf, Akal:2020twv, Akal:2021foz, Kawabata:2021hac, Fernandez-Silvestre:2021ghq, Reyes:2021npy, Akal:2022qei}.

In this paper, we study the properties of entanglement entropy and islands in four-dimensional Schwarzschild black hole following the s-wave approximation, proposed in~\cite{Penington:2019npb} and used in the context of higher-dimensional black holes in~\cite{HIM}. This approximation, having been applied to the fields defined on the background of a higher-dimensional Schwarzschild black hole, effectively reduces the problem to a two-dimensional one. The model~\cite{HIM} explains how the entanglement entropy, associated with semi-infinite regions ``collecting'' Hawking radiation, saturates after taking into account the contribution of the entanglement island in the outer near-horizon zone of two-sided Schwarzschild black hole. The variety of papers exploiting the s-wave approximation in different contexts have been published recently~\cite{Alishahiha:2020qza, Matsuo:2020ypv, Karananas:2020fwx, Yu:2021cgi, Ling:2020laa, Lu:2021gmv, Wang:2021woy, Ahn:2021chg, Arefeva:2021kfx, Yu:2021rfg, Azarnia:2021uch, Cao:2021ujs, Kim:2021gzd, He:2021mst, Ageev:2022hqc, Arefeva:2022cam, Djordjevic:2022qdk, Anand:2022mla, Azarnia:2022kmp, Gan:2022jay}.

We generalize the results of~\cite{HIM} by considering Hawking quanta collected in entangling regions of finite extent. This problem reveals curious features of entanglement entropy in two-sided Schwarzschild black hole. One of the motivations for studying finite regions in asymptotically flat black holes is to simulate a black hole in the spacetime with a positive cosmological constant (Schwarzschild-de Sitter black hole), in which, due to the presence of a cosmological horizon, only a spatial region of finite size is available to the observer.

It is well known~\cite{Solodukhin:2011gn, Nishioka:2018khk} that the entanglement entropy of a pure state is exactly zero, and if this state is bipartite then the entropies of each partition are equal. In the following, we call these properties \textit{pure state condition} and \textit{complementarity}, respectively. The latter is commonly used in the calculations of the entanglement entropy for semi-infinite regions~\cite{HIM, Alishahiha:2020qza, Sybesma:2020fxg, Kim:2021gzd, Goswami:2022ylc}. In Section~\ref{sec:Finite_entangling_regions}, we explicitly check whether these two properties hold for two-sided Schwarzschild spacetime in such a setup, i.e.,
\bea
 \label{eq:PuretyProp}
    & S(\text{pure state}) = 0, \\
    & S(R) = S\left(\overline{R}\right).
    \label{eq:ComplProp}
\eea
We derive a special regularization prescription that allows to preserve complementarity and pure state condition in explicit calculations in two-sided Schwarzschild black hole up to a universal constant term.

In Section~\ref{sec:EntropyFiniteRegions}, we study the time evolution of the entanglement entropy of Hawking radiation for two fundamental types of finite entangling regions, including their island phase.

In Section~\ref{sec:IPFR}, we discuss the information paradox in the context of finite entangling regions.

A brief overview of the setup is given in Section~\ref{sec:Setup}. Section~\ref{sec:ConclusionPerspective} contains a short summary and future prospects. Some technical details have been moved to Appendix~\ref{sec:appendix} for the purpose of readability.

%%%%%%%%%%%%%%%%%%%%%%%%%%%%%%%%%%%%%%%%%%%%%%%%

\section{\label{sec:Setup}Setup}
\subsection{Geometry}

We start with the metric of the four-dimensional Schwarzschild black hole
\be
    ds^2 = -f(r)dt^2 + \frac{dr^2}{f(r)} + r^2 d\Omega_2^2, \qquad f(r) = 1 - \frac{r_h}{r},
    \label{eq:Sch_metric}
\ee
where $r_h$ denotes the black hole horizon, and $d\Omega_2^2$ is the angular part of the metric. Introducing Kruskal coordinates, which for the right wedge take the form
\be
    U = -\frac{1}{\kappa_h}\,e^{-\kappa_h(t - r_*(r))}, \quad V = \frac{1}{\kappa_h}\,e^{\kappa_h(t + r_*(r))},
    \label{eq:right_wedge_Krusk}
\ee
with the tortoise coordinate $r_*(r) = r + r_h\ln\left[(r - r_h)/r_h\right]$ and the surface gravity ${\kappa_h = \nicefrac{1}{2r_h}}$, we can rewrite the metric in the form
\be
    ds^2 = -e^{2 \rho(r)} dU dV + r^2 d\Omega^2,
    \label{eq:krusk-metr}
\ee
with the conformal factor $e^{2 \rho(r)}$ given by
\be
    e^{2 \rho(r)} = \frac{e^{-2\kappa_h r}}{2\kappa_h r}.
    \label{eq:conformalfactor}
\ee

In what follows, we need a formula for the radial distance $d(\bx, \by)$ for the spherically symmetric two-dimensional part of the metric. If we consider~\eqref{eq:krusk-metr} as a Weyl transformed version of the metric ${ds^2 = -dU\,dV}$ (neglecting the angular part) with the Weyl factor $e^{2\rho(r)}$, the square of the distance $d(\bx, \by)$ can be derived as
\be
    d^2(\bx, \by) = e^{\rho(\bx)} e^{\rho(\by)}\left[U(\bx) - U(\by)\right] \left[V(\by) - V(\bx)\right],
    \label{eq:d2}
\ee
where bold letters denote pairs of radial and time coordinates, e.g., $\bx = \left(x, t_x\right)$. In terms of ($t, r$)-coordinates, the distance reads
\be
    \begin{aligned}
        d^2(\bx, \by) & = 
        \frac{2 \sqrt{f(x)f(y)}}{\kappa^2_h} \times \\
        & \times \big[\cosh \kappa_h (r_*(x) - r_*(y)) - \cosh\kappa_h (t_x - t_y)\big].
    \end{aligned}
    \label{eq:geod_dist}
\ee

We use the following notation for spacetime points in the right and left wedges of the Penrose diagram, respectively
\be\nn
    \bx_+ = \left(x_+,\,t_{x_+}\right), \qquad \bx_- = \left(x_-,\,t_{x_-} + \frac{i \pi}{\kappa_h}\right).
    \label{eq:notation_points}
\ee
Note that the imaginary part of the time coordinate of $\bx_-$ implies that this point is in the left wedge.

By the \textit{infrared limit} of the point $\bx$, we mean that $\bx$ tends to spacelike infinity $i^0$ in the corresponding wedge along an arbitrary spacelike curve: $t_x = t_x (x)$.

%----------------------------%

\subsection{Entanglement entropy}

Generally speaking, the calculation of entanglement entropy in a higher-dimensional curved spacetime is an extremely challenging problem. An important suggestion made in~\cite{HIM} was to consider the s-wave approximation. The fact that a static observer at spatial infinity collects predominantly lower multipoles, while the higher ones backscatter in the Schwarzschild black hole potential, reduces the initially complicated setup to a simpler two-dimensional problem of calculation the entanglement entropy of conformal matter. In this case, the following expression of the entanglement entropy for $N \geq 1$ separate intervals is used
\be
    \begin{aligned}
        S_\m = &\frac{c}{3} \sum_{i,\,j} \ln\frac{d(\bx_{i}, \by_{j})}{\eps} - \\
        - &\frac{c}{3} \sum_{i\,<\,j}^{N}\ln\frac{d(\bx_{i}, \bx_{j})}{\eps} - \frac{c}{3} \sum_{i\,<\,j} \ln\frac{d(\by_{i}, \by_{j})}{\eps},
    \end{aligned}
    \label{eq:S_N_ints}
\ee
where the distance $d(\bx_i,\by_j)$ is given by~\eqref{eq:geod_dist}, $\bx_i$ and $\by_i$ denote left and right endpoints of the corresponding intervals, and $\eps$ is a UV cutoff. 

Few remarks are in order. First, one can implicitly assume that this formula describes the entanglement entropy of $c$ free massless Dirac fermions. For flat spacetime, this formula for Dirac fermions was obtained in~\cite{Casini:2005rm}. Taking into account the transformation properties of entanglement entropy under Weyl transformations~\cite{Almheiri:2019psf}, one can assume the validity of~\eqref{eq:S_N_ints} for the curved background~\eqref{eq:krusk-metr} at fixed values of angular coordinates. The entanglement entropy of free massless Dirac fermions in a curved background has been considered in the context of the island proposal~\cite{Almheiri:2019qdq,Azarnia:2021uch, Kawabata:2021vyo} and for inhomogeneous non-interacting Fermi gases \cite{Dubail:2016tsc}.

Second, there are two versions of the theory of free Dirac fermions, which differ in whether the fermion number is $\mathbb{Z}_2$-gauged or not. As it was shown in~\cite{Headrick:2012fk}, if the number of fermions is not gauged, then there is no modular invariance in the theory, and vice versa \footnote{We thank an anonymous referee for pointing out to this issue.}. At the same time, the derivation of the formula~\eqref{eq:S_N_ints} in flat spacetime~\cite{Casini:2005rm} is based on ungauged theory. Modular transformations are ``large'' diffeomorphisms, and the absence of modular invariance could be an obstacle in obtaining the formula~\eqref{eq:S_N_ints} for free Dirac fermions in a curved background. However, there are two special cases. For one interval, $N = 1$, the formula~\eqref{eq:S_N_ints} coincides with the entanglement entropy of any  two-dimensional CFT. For $N = 2$ intervals, the entanglement entropy in the modular invariant theory of two-dimensional free compact boson at the self-dual radius coincides with the entanglement entropy~\eqref{eq:S_N_ints} of free Dirac fermions~\cite{Calabrese:2009ez}. These facts give an independent support to the formula~\eqref{eq:S_N_ints} for $N = 1$ and $N = 2$ intervals.

\subsection{Generalized entropy functional}

Recently, it was shown that the expected behavior of the Page curve emerges from the island proposal~\cite{Cotler:2017erl, Almheiri:2019hni, Penington:2019npb, Almheiri:2019qdq}. Considering the Hartle-Hawking vacuum~\cite{Hartle:1983ai}, the reduced density matrix of Hawking radiation collected in~$R$ is defined by tracing out the states in the complement region~$\overline{R}$, which includes the black hole interior. The island mechanism prescribes that the states in some regions~${I \subset \overline{R}}$, called entanglement islands, are to be excluded from tracing out.

The island contribution can be taken into account via the generalized entropy functional defined as~\cite{Penington:2019kki, Almheiri:2019qdq}
\be
    S_\gen[I, R] = \frac{\operatorname{Area}(\partial I)}{4\GN} + S_\m(R \cup I).
    \label{eq:gen_functional}
\ee
Here $\partial I$ denotes the boundary of the entanglement island, $\GN$ is Newton's constant, and $S_\m$ is the entanglement entropy of conformal matter. One should extremize this functional over all possible island configurations
\be
    S\extgen[I, R] = \underset{\partial I}{\operatorname{ext}}\,\Big\{S_\gen[I, R]\Big\},
\ee
and then choose the minimal one
\be
    S(R) = \underset{\partial I}{\text{min}}\,\Big\{S\extgen[I, R]\Big\}.
    \label{eq:Sgen}
\ee

%%%%%%%%%%%%%%%%%%%%%%%%%%%%%%%%%%%

\section{\label{sec:Finite_entangling_regions}Infrared regularization of entanglement entropy}

In this section, we consider different partitions of Cauchy surfaces in Schwarzschild spacetime and propose a regularization of spacelike infinities, which allows to preserve complementarity and pure state condition of entanglement entropy within the framework of the formula~\eqref{eq:S_N_ints}.
\skipline

\begin{figure}[h]\centering
    \includegraphics[width=0.5\textwidth]{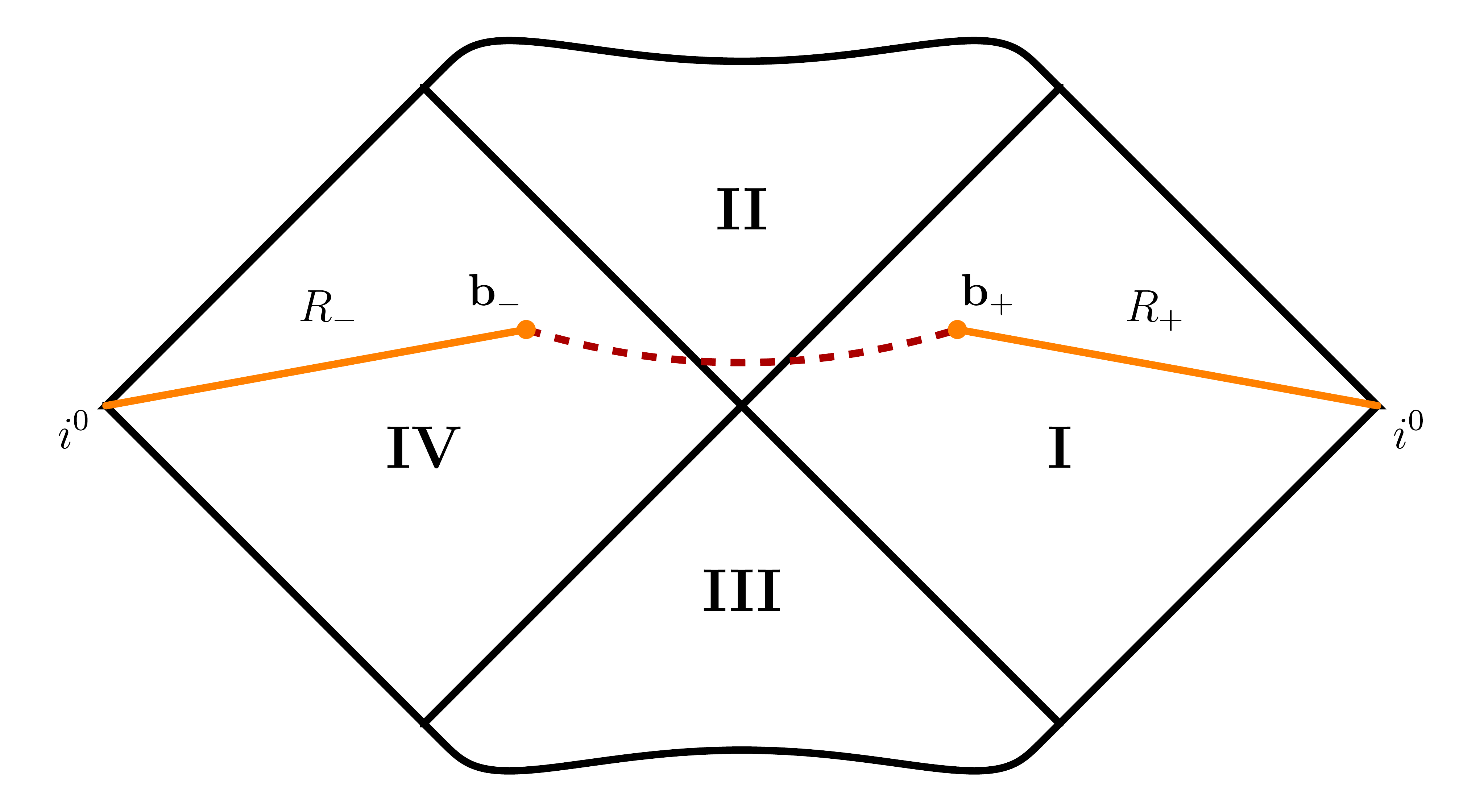}
	\caption{Penrose diagram for the eternal Schwarzschild black hole with the \textit{schematic} plots of the entangling region ${R_\infty \equiv R_- \cup R_+}$~(orange) and its complement ${\overline{R}_\infty \equiv [\bb_-,\,\bb_+]}$~(dashed dark red).}
    \label{fig:setup-HIM-NoIsl}
\end{figure}

The issue can be seen from the following reasoning. Previous papers (see, for example,~\cite{HIM, Alishahiha:2020qza, Sybesma:2020fxg, Kim:2021gzd, Goswami:2022ylc}) have extensively used the calculation of the entropy for finite complements of semi-infinite entangling regions. In particular, the entanglement entropy for the semi-infinite region ${R_\infty \equiv R_- \cup R_+ = (i^0,\,\bb_-] \cup [\bb_+,\,i^0)}$, where the Hawking radiation is collected, is actually calculated in~\cite{HIM} using the complementarity property
\be
    S_\m(R_\infty) = S_\m\left({\bar R_\infty}\right),
\ee
with the complement $\overline{R}_\infty = [\bb_-,\,\bb_+]$ (see Fig.~\ref{fig:setup-HIM-NoIsl}). The calculation of $S_\m\left(\overline{R}_\infty\right)$ leads to~\cite{HIM}
\be\label{eq:without_island}
    S_\m\left(\overline{R}_\infty\right) = \frac{c}{6}\ln\left(\frac{4f(b)}{\kappa_h^2\eps^2}\cosh^2\kappa_h t_b\right).
\ee
In the limit when the boundaries of the entangling regions are sent to spacelike infinities $i^0$ along the same timeslices (i.e., $b \to \infty$ at fixed $t_b$), this formula reduces to
\be
    \lim\limits_{b \to \infty} S_\m\left(\overline{R}_\infty\right) = \frac{c}{3}\ln\frac{2}{\kappa_h\eps} + \frac{c}{3}\ln\cosh\kappa_h t_b.
    \label{eq:HIM_limit}
\ee
This is not what to be expected, since in this limit of the vanishingly small region $R_\infty$, entangled particles of radiation are not collected, and the entanglement entropy~\eqref{eq:without_island} should have been equal to zero. However, the result grows linearly at late times. This raises a question on whether it is possible to introduce a prescription, which allows to calculate the entropy for semi-infinite entangling regions directly.

%----------------------------%

\subsection{\label{sec:warm-up}Warm-up: complementarity and pure state condition in CFT in flat background}

Let us consider a finite interval $R = \left[-\ell/2, \, \ell/2\right]$ on the plane $\mathbb{R}^2$ with coordinates $(\tau, x)$ at a constant time $\tau = 0$. It is well known \cite{Calabrese:2004eu, Calabrese:2009qy} that the entanglement entropy for the subsystem $R$ is
\be\label{eq:basiqcft2}
    S_\m(R) = \frac{c}{3}\ln\frac{\ell}{\eps}.
\ee
Suppose that we are given a pure state on the entire line $\tau = 0$, which is a hypersurface $\mathbb{R}^1$. Pure state condition~\eqref{eq:PuretyProp} requires $S_\m\left({\mathbb R}^1\right) = 0$. Sending $\ell \to \infty$, we obtain ${\mathbb R}^1$, so we should have had $\lim_{\ell\to\infty}S_{\text{m}}(R) \to 0$. However, the entropy as given by~\eqref{eq:basiqcft2} obviously diverges as $\ln \ell$ when $\ell \to \infty$. This simply means that the equation~\eqref{eq:basiqcft2} does not hold for large $\ell$. In this regard, one has to specify the size of the system, say $L$, and then send $\ell$ to $L$. 

\begin{figure}[h]\centering
    \includegraphics[width=0.25\textwidth]{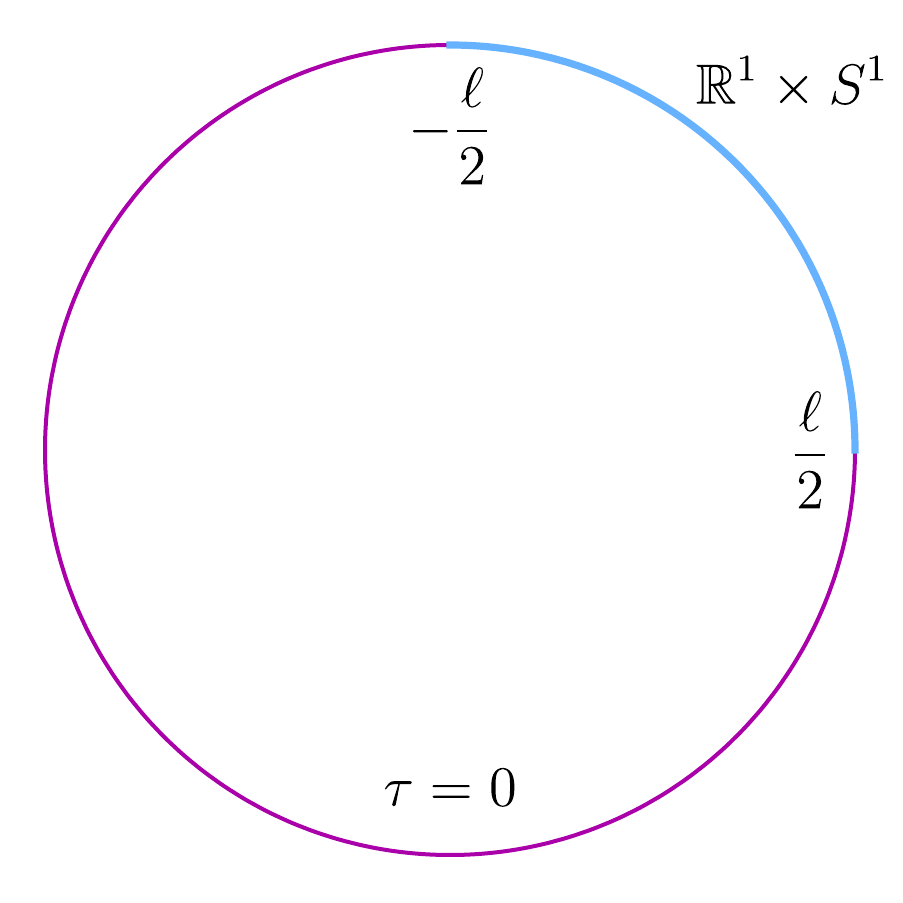}
    \caption{The interval $\left[-\ell/2,\,\ell/2\right]$ (blue) and its complement (magenta) on the line $\tau = 0$ of the cylinder $\mathbb{R}^1 \times S^1$ with circumference $L$.}
    \label{fig:CFT-Flat-Figs2}
\end{figure}

A possible way to deal with a system of a finite size $L$ is to consider a cylinder with circumference $L$ (see Fig.~\ref{fig:CFT-Flat-Figs2}). In this case, the entanglement entropy for an interval of length~$\ell$ is given by
\be\label{eq:cft2circle}
    S_\m(R) = \frac{c}{3}\ln\left(\frac{L}{\pi\eps}\sin\frac{\pi\ell}{L}\right).
\ee
In the limit $L \to \infty$ with $\ell$ kept fixed, the entanglement entropy~\eqref{eq:basiqcft2} for an interval on the plane $\mathbb{R}^2$ is reproduced. Due to the symmetry $\ell \to L - \ell$ of the expression~\eqref{eq:cft2circle}, the complementarity holds automatically. Moreover, when we lengthen the interval $R$ such that the whole system is considered, i.e.,~$\ell \to L$ (in fact, $\ell \to L - \eps$ for the system on a lattice), the entanglement entropy~\eqref{eq:cft2circle} becomes 
\be\nn
    S_\m(R) = \frac{c}{3}\lim_{\varepsilon\to 0}\ln\left(\frac{L }{\pi\varepsilon }\sin \left(\frac{\pi (L-\varepsilon )}{L}\right)\right) = 0. 
\ee
These simple considerations show that by introducing a suitable regularization we can preserve purity state condition and complementary property in this approximation. 

An analogous closure of hypersurfaces of constant time, such that it turns into $S^1$-topology, is impossible in eternal Schwarzschild black hole without violating the causal structure of this spacetime, since the left and right wedges are causally disconnected. However, we can refer to other properties of Schwarzschild spacetime. Note that there are conformal factors in the formula for entanglement entropy in a curved background, see~\eqref{eq:S_N_ints} and~\eqref{eq:d2}. In Schwarzschild geometry, these conformal factors in Kruskal coordinates are given by~\eqref{eq:conformalfactor} and tend to zero at $r \to \infty$. In addition to the fact that there are two spacelike infinities $i^0$ in this spacetime, one can, in principle, argue that it is possible to obtain a finite distance between spacelike infinities in the left and right wedges after mutual cancellation of IR divergences without imposing periodic boundary conditions. 
In what follows, we will show that we can explicitly preserve complementarity and pure state condition of entanglement entropy with its approximate expression for multiple intervals in eternal Schwarzschild black hole with one (two) endpoints(s) going to spacelike infinity~$i^0$ in a special way, rather than cut off semi-infinite intervals.

%----------------------------%

\subsection{\label{sec:IR-Cauchy}Infrared regularization of Cauchy surfaces}

\begin{figure}[h]\centering
    \includegraphics[width=0.5\textwidth]{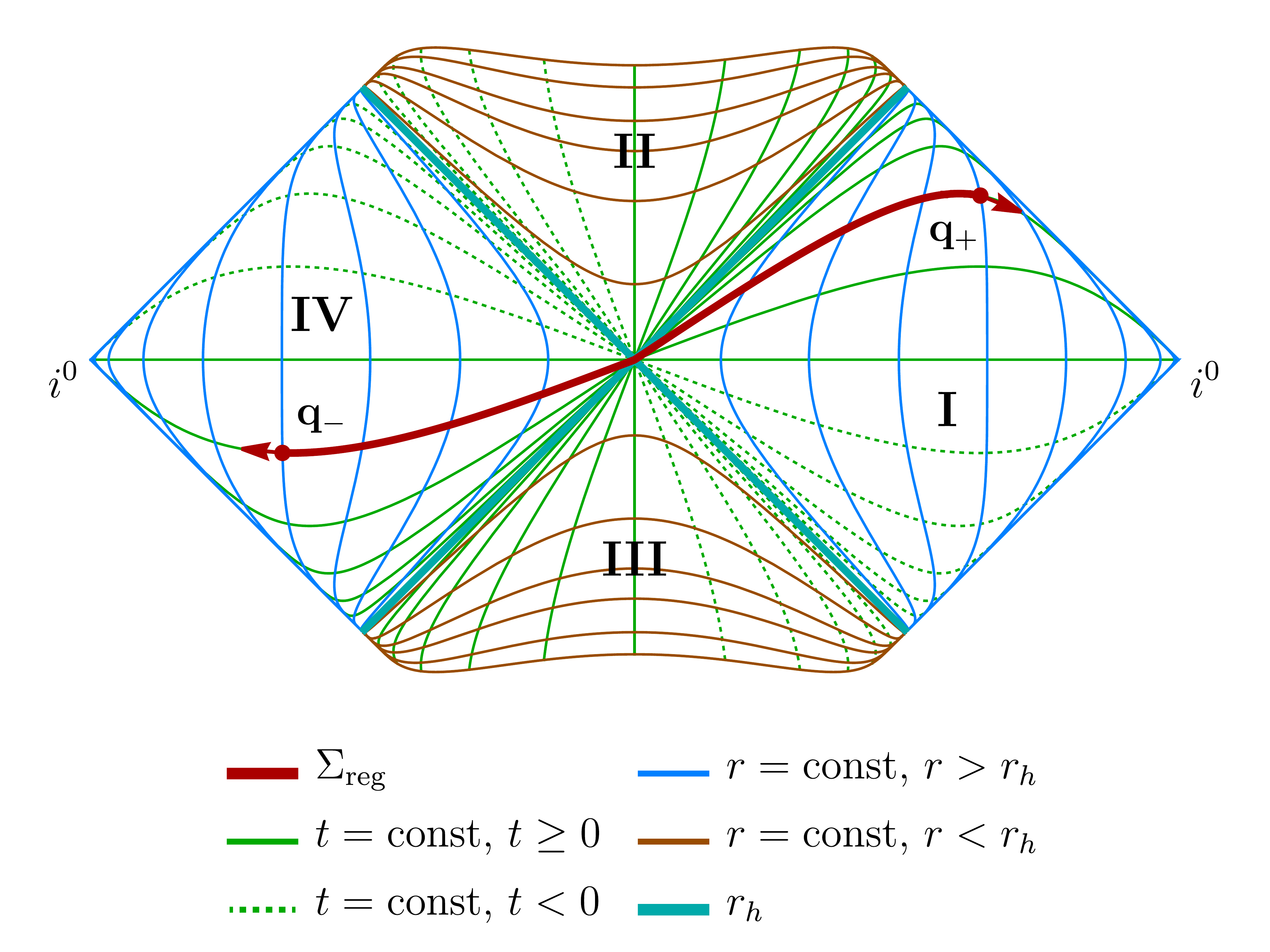}
	\caption{Penrose diagram for the eternal Schwarzschild black hole with the regularized Cauchy surface $\Sigma_\text{reg} \equiv [\bq_-,\,\bq_+]$. Arrows highlight that the regulators tend to spacelike infinities, $\bq_\pm \to i^0$.}
    \label{fig:CauchySurf}
\end{figure}

Let us introduce an IR regularization of a Cauchy surface $\Sigma$ which extends between spacelike infinities in the left and right wedges of Schwarzschild (see Penrose diagram on Fig.~\ref{fig:CauchySurf}). Since the Hartle--Hawking state given on $\Sigma$ is pure, we should have $S_\m(\Sigma) = 0$ for the entanglement entropy. To regularize $\Sigma$, we take a finite spacelike interval $\Sigma_\text{reg} \subset \Sigma$ such that ${\Sigma_\text{reg} = [\bq_-,\,\bq_+]}$ with the following coordinates of the IR regulators $\bq_\pm$
\be\nn
    \bq_+ = \Big(q_+,\,t_{q_+}(q_+)\Big), \qquad \bq_- = \left(q_-,\,t_{q_-}(q_-) + \frac{i\pi}{\kappa_h}\right).
\ee

Then, the entanglement entropy of conformal matter on the regularized Cauchy surface $\Sigma_\text{reg}$ is given by
\be
    \begin{aligned}
        & S_\m\left(\Sigma_\text{reg}\right) = \frac{c}{3}\ln\frac{d(\bq_-,\,\bq_+)}{\eps} = \\
        & = \frac{c}{6}\ln\left[\frac{2 \sqrt{f(q_+) f(q_-)}}{\kappa^2_h\eps^2}\Big(\cosh\kappa_h (r_*(q_+) - r_*(q_-))\right. + \\
        & + \left.\cosh\kappa_h (t_{q_+}(q_+) - t_{q_-}(q_-))\Big)\right],
    \end{aligned}
    \label{eq:reg-Cauchy-gen}
\ee
which is, in general, divergent in the IR limit $q_{\pm} \to \infty$. However, this limit exists along the curves $\mathcal{C}$ of the following type
\be
    \mathcal{C}:\quad
        \begin{aligned}
            & \kappa_h\left(r_*(q_+) - r_*(q_-)\right) = c_1 + \alpha (q_+, q_-), \\
            & \kappa_h\left(t_{q_+} (q_+) - t_{q_-} (q_-)\right) = c_2 + \beta (q_+, q_-),
        \end{aligned}
    \label{eq:curve}
\ee
where $c_1$ and $c_2$ are arbitrary constants fixed for a particular curve, and
\be\nn
    \lim_{q_{\pm} \to \infty} \alpha (q_+, q_-) = 0, \qquad \lim_{q_{\pm} \to \infty} \beta (q_+, q_-) = 0.
\ee

Taking into account that $f\left(q_{\pm}\right) \to 1$ as $q_{\pm} \to~\infty$, the IR limit of~\eqref{eq:reg-Cauchy-gen} along some curve $\mathcal{C}$~\eqref{eq:curve} is given by
\be\label{eq:cauchyundercurve}
    S_\m(\Sigma) = \lim\limits_{\substack{q_\pm \to \infty \\ \bq_{\pm} \in \mathcal{C}}} S_\m\left(\Sigma_\text{reg}\right) = \frac{c}{3}\ln \frac{2}{\kappa_h \eps} + F\left(c_1, c_2\right),
\ee
where
\be
    F\left(c_1, c_2\right) \equiv \frac{c}{6}\ln\left(\frac{\cosh c_1 + \cosh c_2}{2}\right) \geq 0.
    \label{eq:F}
\ee
Since this function is non-negative, the entanglement entropy can take any positive value by varying the constants $c_1$, $c_2$. For the same reason, we cannot get rid of the first term on the RHS of~\eqref{eq:cauchyundercurve} by an appropriate choice of the constants $c_1$, $c_2$. The best we can get is to let $c_1 = c_2 = 0$, which according to~\eqref{eq:curve} leads to
\be\nn
    \lim\limits_{\substack{q_\pm \to \infty \\ \bq_{\pm} \in \mathcal{C}}}\left(q_+ - q_-\right) = 0, \quad \lim\limits_{\substack{q_\pm \to \infty \\ \bq_{\pm} \in \mathcal{C}}}\Big(t_{q_+}(q_+) - t_{q_-}(q_-)\Big) = 0.
\ee
This means that the IR limit is taken to be asymptotically radially symmetric, and the regulators, as they go to spacelike infinity, asymptotically approach the same timeslice. For this case, the entropy of the Hartle-Hawking state defined on~$\Sigma$ is given by
\be
    S_\m(\Sigma)\Big|_{c_1\,=\,c_2\,=\,0} = \frac{c}{3}\ln \frac{2}{\kappa_h\eps}.
    \label{eq:IR_anomaly}
\ee
Since the entropy of a pure state should be zero, we claim that this anomalous term is to be subtracted from final answers.

At first sight, it might seem surprising that the result in the IR limit~\eqref{eq:IR_anomaly} depends on the UV cutoff $\eps$. In fact, this is to be expected, since~$\eps$ is the only dimensional constant, in addition to $r_h$, to make the whole answer dimensionless.

The reasons why we get the entanglement entropy for infinite intervals that is not infrared divergent are the special behavior of the Weyl factor~\eqref{eq:conformalfactor} at infi\-nity, and the mutual asymptotic cancellation of the regulators. The presence of two regulators of the same sign, which cancel each other in final results, is the hallmark of two-sidedness and higher dimensionality\footnote{In fact, there are two different static patches $(t,r)$ in two-sided Schwarzschild --- in the left and right wedges, respectively. Spacelike infinities $r \to \infty$ are positive due to the positiveness of the radial coordinate $r\geq 0$, which can be introduced for spacetimes of higher dimensionality. For example, this is not applicable for two-dimensional Minkowski spacetime, in which spatial coordinate $x \in (-\infty, \infty)$. Therefore, there are two spacelike infinities of different signs: $x \to \pm \infty$. This leads to the IR divergent entanglement entropy for the Cauchy surface: $S \propto \ln L$ as $L \to \infty$.} of the eternal Schwarzschild black hole geometry, and cannot be seen as a general method of regularization of infinities during the calculation of entanglement entropy for infinite regions.

%-------------------------------%

\subsection{Infrared regularization of semi-infinite complement of finite entangling regions}

\begin{figure}[h]\centering
    \includegraphics[width=0.5\textwidth]{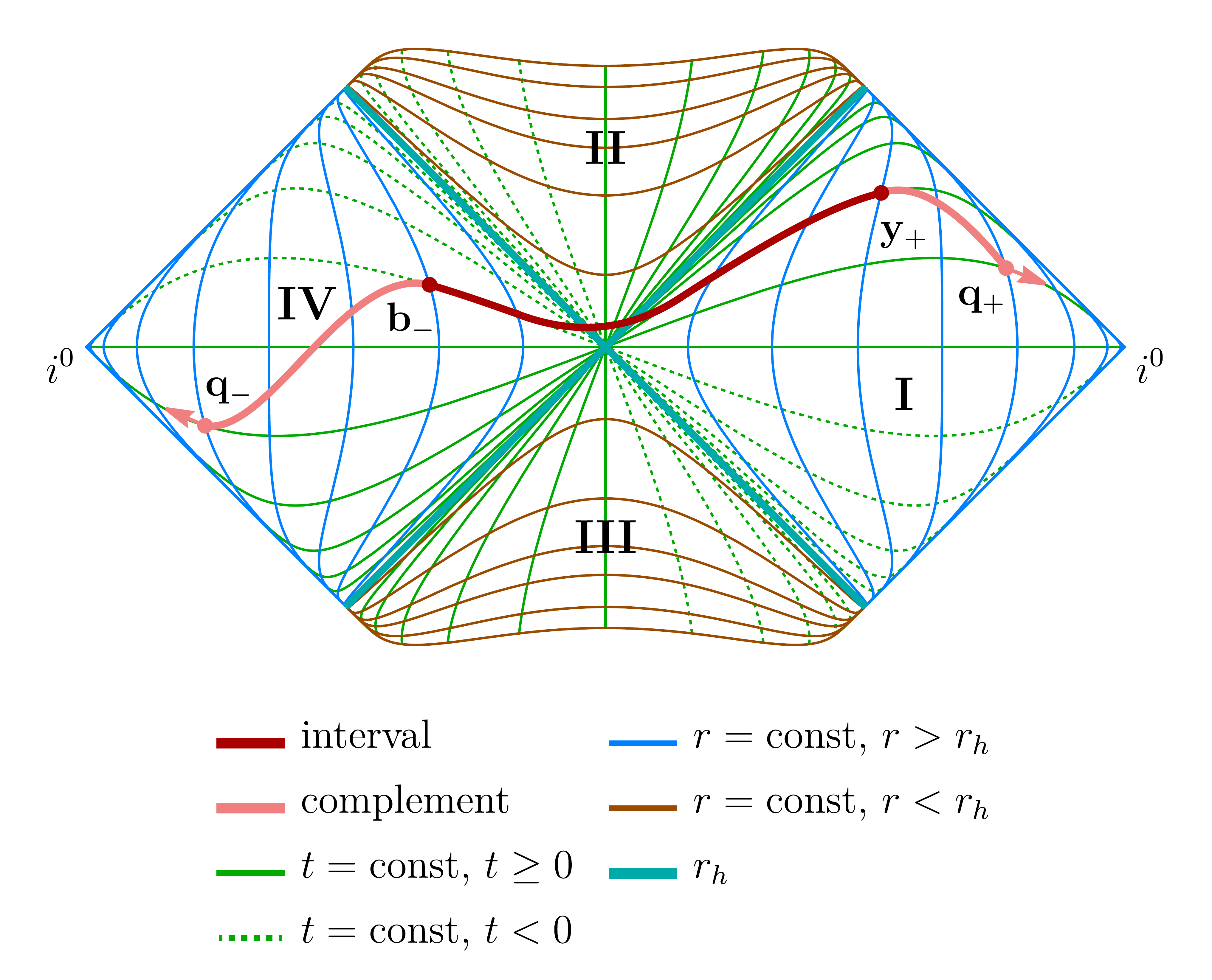}
    \caption{Penrose diagram for the eternal Schwarzschild black hole with the regularized finite interval configuration: the finite interval~$[\bb_-,\,\by_+]$ (dark red) and its regularized complement~$[\bq_-,\,\bb_-] \cup [\by_+,\,\bq_+]$ (light red). Arrows mean that the regulators tend to spacelike infinities, $\bq_\pm \to i^0$.}
    \label{fig:CheckComplPropCase2}
\end{figure}

A Cauchy surface can be divided into any number of finite entangling regions. A single finite region can lie entirely in one wedge or be extended over two. Regardless, the complement $\overline{R}$ includes two semi-infinite intervals in each wedge, which stretch to the corresponding spacelike infinities $i^0$. Our goal is to represent the complement as the IR limit of some finite regions (see Fig.~\ref{fig:CheckComplPropCase2}). One can consider this procedure as a regularization of spacelike infinities. In this way, the endpoints are defined as
\be\nn
    \begin{aligned}
        & \by_+ = \left(y,\,t_y\right), \quad \bb_+ = \left(b,\,t_b\right), \quad \bb_- = \left(b,\,t_b + \frac{i \pi}{\kappa_h}\right), \\
        & \bq_+ = \Big(q_+,\,t_{q_+}(q_+)\Big), \qquad \bq_- = \left(q_-,\,t_{q_-}(q_-) + \frac{i \pi}{\kappa_h}\right).
    \end{aligned}
\ee

Now we want to check explicitly the complementarity property $S(R) = S\left(\overline{R}\right)$ for the region $R$ and its complement $\overline{R}$. In gravitational theories, both $S(R)$ and $S\left(\overline{R}\right)$ are determined by the island formula~\eqref{eq:Sgen}, which contains the ``area term'' and the matter contribution~$S_\m$. If the island $I$ is the same both for $R$ and $\overline{R}$, to verify $S(R) = S\left(\overline{R}\right)$ it suffices to compare the entanglement entropies of matter only, $S_\m (R \cup I) = S_\m \left(\overline{R \cup I}\right)$. Indeed, in this case, the island formulas for the region $R$ and the complement $\overline{R}$ yield the same, because the same expression is extremized in~\eqref{eq:Sgen}. Therefore, further in this section, we consider only the entanglement entropy of matter $S_\m$.

The entanglement entropy for the region ${R = [\bb_\pm,\,\by_+]}$ reads
\be
    S_\m(R) = \frac{c}{3}\ln\frac{d(\bb_\pm, \by_+)}{\eps}.
\ee 
We compare this with the entanglement entropy for the regularized complement $\overline{R}_\text{reg} = [\bq_-,\,\bb_\pm] \cup [\by_+,\,\bq_+]$
\be
    \begin{aligned}
        S_\m\left(\overline{R}_\text{reg}\right) & = \frac{c}{3}\ln\frac{d(\bb_\pm, \by_+)}{\eps} + \frac{c}{3}\ln\frac{d(\bq_-, \bq_+)}{\eps} + \\
        & + \frac{c}{3}\ln\left[\frac{d(\by_+, \bq_+)d(\bb_\pm, \bq_-)}{d(\by_+, \bq_-)d(\bb_\pm, \bq_+)}\right].
    \end{aligned}
\ee
The third term on the RHS tends to zero in the IR limit taken along any spacelike curve. This is because the distance to the regulator $\bq_+$ in the numerator has a counterpart in the denominator, and both have similar asymptotic behavior. The same holds for the distances to the regulator $\bq_-$. Due to the generically divergent distance between the regulators encountered in the second term on the RHS, the IR limit of this expression exists only along the curves~ given by $\mathcal{C}$~\eqref{eq:curve} and yields
\be 
    \lim\limits_{\substack{q_\pm \to \infty \\ \bq_{\pm} \in \mathcal{C}}} S_\m\left(\overline{R}_\text{reg}\right) = S_\m(R) + \frac{c}{3}\ln\frac{2}{\kappa_h \eps} + F\left(c_1, c_2\right),
    \label{eq:Rr-R}
\ee
This result does \textit{not} depend on the wedge (left, right or both) in which the finite entangling region is located. The last two terms of this expression are exactly the same as we get when regularizing a Cauchy surface~\eqref{eq:cauchyundercurve}.
\skipline

It is straightforward to extend this calculation to the case when $R$ consists of $N$ disjoint intervals: ${R = [\ba_1, \, \ba_2] \cup \ldots \cup [\ba_{2N - 1}, \, \ba_{2N}]}$. In this case, the entanglement entropy for the regularized complement is given~by
\be
    \begin{aligned}
        S_\m\left(\overline{R}_\text{reg}\right) & = S_\m(R) + \frac{c}{3}\ln\frac{d(\bq_-,\bq_+)}{\eps} + \\
        & + \frac{c}{3}\ln\left[\frac{d(\ba_1,\bq_-) \ldots d(\ba_{2N},\bq_+)}{d(\ba_1,\bq_+) \ldots d(\ba_{2N},\bq_-)}\right].
    \end{aligned}
    \label{eq:Nint}
\ee
The third term on the RHS, which makes the only difference with the case of a single interval, is vanishingly small in the IR limit along spacelike curves. Along the curves $\mathcal{C}$~\eqref{eq:curve}, the whole expression becomes the same as~\eqref{eq:Rr-R}.

%----------------------------%

\subsection{\label{sec:EqualTimeSlice}Infrared regularization consistent with complementarity and pure state condition}

Let us sum up the results. We have established that in direct calculations, complementarity and pure state condition are not complied due to the anomalous term~\eqref{eq:IR_anomaly} and the function $F(c_1, c_2)$~\eqref{eq:F}, which depends on arbitrary parameters $c_1$ and $c_2$.

Note that the violation of pure state condition and complementarity may be related to our approach to IR regularization, the shortcomings of which we discussed for the flat case in~\ref{sec:warm-up}. However, due to the features of the analytically extended Schwarzschild geometry, we obtained a violation up to a constant, and not up to an IR divergent term, as in flat spacetime.

The IR limit of the entropy is well-defined along the special class of curves $\mathcal{C}$~\eqref{eq:curve} and for any choice of the constants $c_1$ and $c_2$. Since $F(c_1,c_2) \geq 0$, we let $c_1 = c_2 = 0$, for which $F(0,0) = 0$, to get rid from this contribution. This means that the regulators are to be sent to infinity asymptotically radially symmetric, and as they approach $i^0$, they asymptotically fall on the same timeslice.

Then we \textit{prescribe} that we should subtract the anomalous term \eqref{eq:IR_anomaly}. In the end, we obtain the entanglement entropy for semi-infinite regions consistent with complementarity and pure state condition. 
\skipline

%%%%%%%%%%%%%%%%%%%%%%%%%%%%%%%%%%%%%%%%%%%%%%%%

\section{\label{sec:EntropyFiniteRegions}Entropy dynamics for finite entangling regions}

\begin{figure}[h!]\centering
    \includegraphics[width=0.5\textwidth]{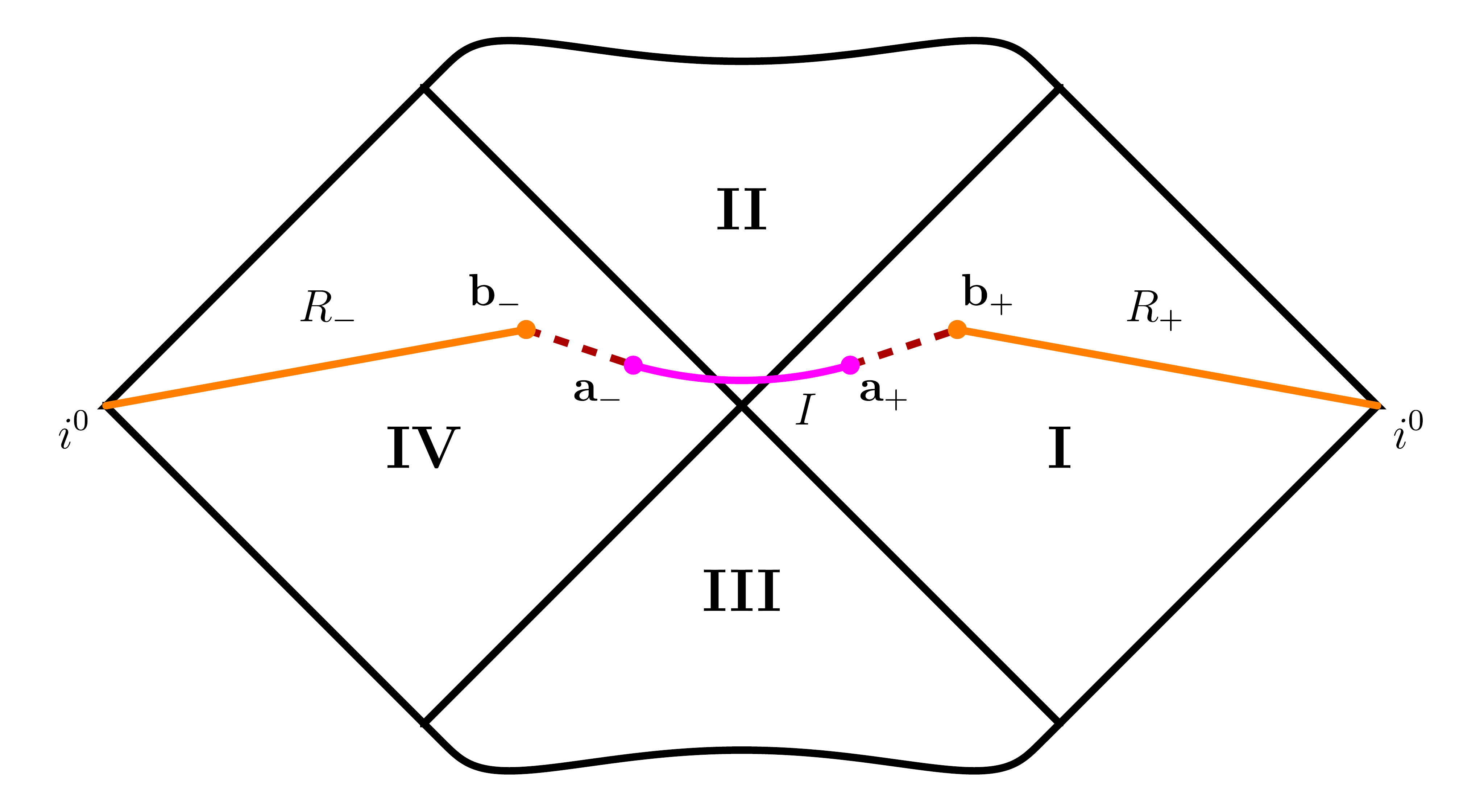}
	\caption{Penrose diagram for the eternal Schwarzschild black hole with the \textit{schematic} plots of semi-infinite entangling region ${R_\infty \equiv R_- \cup R_+}$ (orange), the entanglement island ${I \equiv [\ba_-,\,\ba_+]}$ (magenta) and their complement ${\overline{R_- \cup R_+ \cup I} \equiv [\bb_-,\,\ba_-] \cup [\ba_+,\,\bb_+]}$ (dashed dark red).}
    \label{fig:setup-HIM}
\end{figure}

In this section, we study the dynamics of finite \mbox{entangling} regions affected by the presence of entanglement islands.
\skipline

In the original setup~\cite{HIM}, which describes semi-infinite entangling region $R_\infty = (i^0,\,\bb_-] \cup [\bb_+\,i^0)$, the entanglement entropy~\eqref{eq:without_island} enters an unbounded linear growth regime at late times $t_b \gg r_h$
\be
    S_\m\left(R_\infty\right) \simeq \frac{c}{3}\,\kappa_h t_b.
    \label{eq:late_time_without_island}
\ee
This can be interpreted as a version of the information paradox. It was shown there that taking into account the contribution of the entanglement island leads to saturation of the entanglement entropy~$S(R_\infty)$. At late times, the island is symmetric and its endpoints are located in different wedges near the black hole horizon (see Fig.~\ref{fig:setup-HIM}). The entanglement entropy in the leading-order expansion in $c\,\GN/r^2_h$ reads
\be
    S(R_\infty)\LateTimes \simeq \frac{2\pi r_h^2}{\GN} + \frac{c}{6}\ln\frac{f(b)}{\kappa^4_h\eps^4} + \frac{c}{6}\left(2\kappa_h r_*(b) - 1\right),
    \label{eq:late_time_with_island}
\ee
which is constant in time. We emphasize that this result of~\cite{HIM} fully relies on the complementarity property.

%----------------------------%

\subsection{\label{sec:pick-Cauchy} Geometry and dynamics of Cauchy surfaces}

\subsubsection*{Up-down notation}

We introduce two indices for spacetime points, ``up'' and ``down'', meaning their location with respect to the horizontal line $t = 0$, which stretches between spacelike infinities in the left and right wedges. This notation will prove convenient below. In the following, we use the points
\be
    \begin{aligned}
        & \bb^{\up}_+ = \left(b,\, t_b\right), \quad \bb^{\up}_- = \left(b,\, -t_b + \frac{i\pi}{\kappa_h}\right), \\
        & \bq^{\up}_+ = \left(q,\, t_b\right), \quad \bq^{\up/\down}_- = \left(q,\,\mp t_b + \frac{i\pi}{\kappa_h}\right).
    \end{aligned}
    \label{eq:up-down}
\ee
Note that points $\bb_+$ and $\bb_-$ have the same radial coordinates. Real parts of their time coordinates are opposite in sign, such a choice makes the problem time-dependent~\cite{Almheiri:2019yqk, HIM}. Also, without loss of generality, we take $\bq_{\pm}^{\up/\down}$ radially symmetric.

\subsubsection*{Dynamics of Cauchy surfaces}

Given $N > 2$ points on the Penrose diagram, we can stretch a hypersurface to all of them as well as to the corresponding spacelike infinities $i^0$. Here we develop a recipe for how to select these points so that the resulting hypersurface is a Cauchy one.
\skipline

Let us divide all the points of the regularized Cauchy surface into the inner ones, whose radial coordinates are fixed and finite, and IR regulators $\bq_\pm$. The latter are the endpoints of the regularized finite hypersurface, which are to be sent to $i^0$.

In the following, we discuss the evolution of finite regions of the type $R = [\bq_-^{\up/\down},\,\bb_-^\up] \cup [\bb_+^\up,\,\bq_+^\up]$ (the intermediate points $\bq_+^\up$ and $\bq_-^{\up/\down}$, whose radial coordinates are fixed, should not be confused with the IR regulators $\bq_\pm$). Such a choice of the region endpoints is motivated by the following reason. Moving the intermediate points along their Killing vectors $\partial_t^+$ in the right wedge and $\partial_t^-$ in the left is an isometry of the Schwarzschild spacetime. Thus, we would not get any non-trivial dynamics out of their flow. On the contrary, if we explicitly impose the relations~\eqref{eq:up-down}, we would get the points in the right wedge moving along their Killing vectors, while in the left edge, those points with $t < 0$ would move upwards to $t \to -\infty$, and with $t > 0$~--- downwards to $t \to +\infty$. This choice is \textit{not} a symmetry of the problem, therefore, it would generate a non-trivial dynamics of the entropy.~\cite{Almheiri:2019yqk, Hartman:2013qma}

\begin{figure}[h]\centering
    \includegraphics[width=0.5\textwidth]{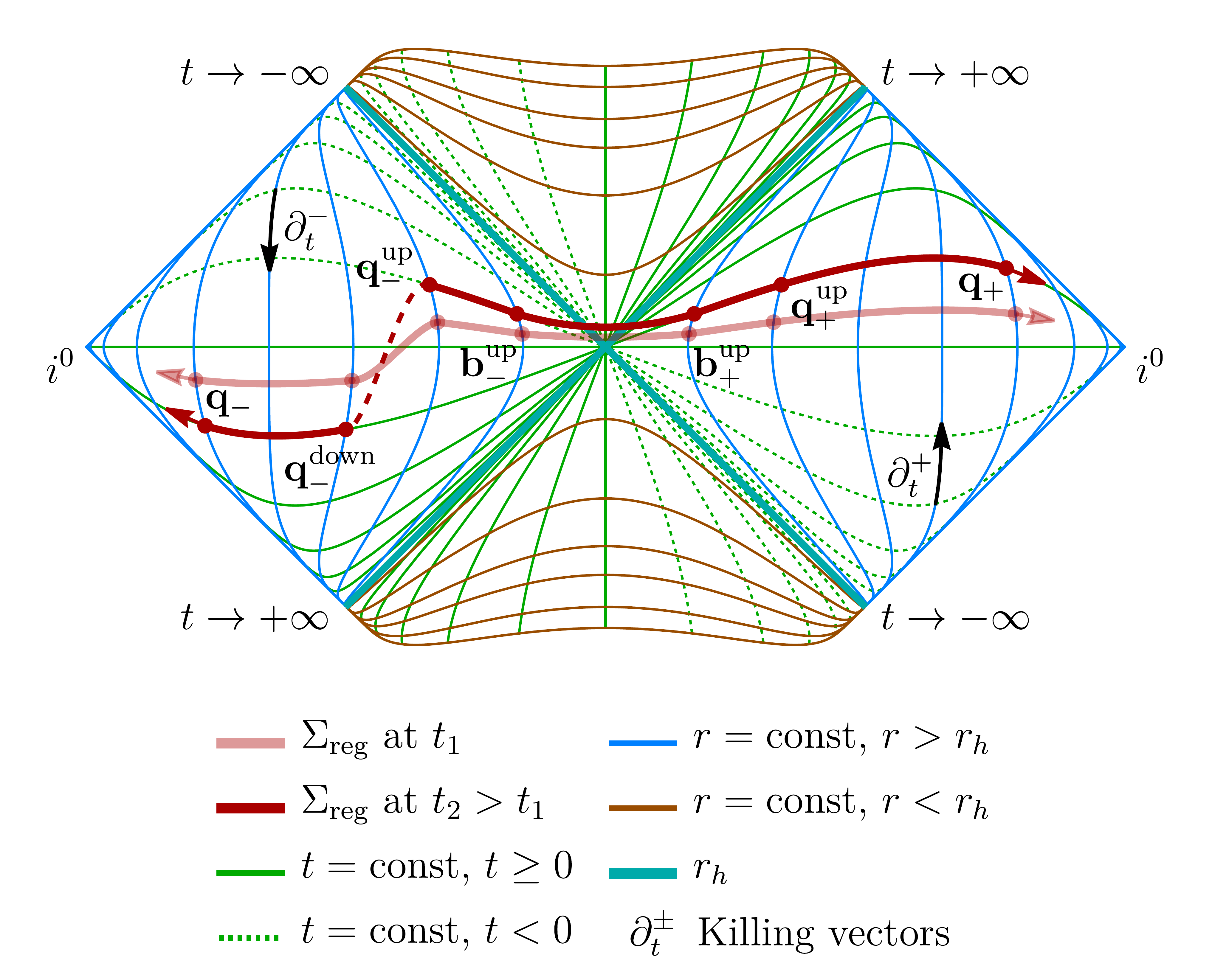}
    \caption{Penrose diagram for the eternal Schwarzschild black hole with a regularized Cauchy surface. During time evolution (from the light red curve to the dark red one) two adjacent intermediate points in the left wedge become timelike separated which breaks the Cauchy surface (dashes).}
    \label{fig:CauchySurfDynamics}
\end{figure}

Having said this, we should emphasize that in the context of finite size regions, the choice of the points~\eqref{eq:up-down} has a non-trivial consequence which we call the ``Cauchy surface breaking''. This phenomenon takes place when some points get separated during time evolution by a timelike interval (see Fig.~\ref{fig:CauchySurfDynamics}) due to different directions of their movement on the diagram. When the Cauchy surface breaks, the problem of studying the evolution of entanglement entropy becomes ill-defined, so we should keep track of this phenomenon.

Thus, the recipe of picking a Cauchy surface is as follows:
\begin{itemize}
    \item[a)] Since a Cauchy surface is a spacelike hypersurface, all its tangents should be spacelike. For our setup this means that if some of the inner points of a hypersurface in the left wedge lie on different sides with respect to the line $t = 0$, they will eventually become timelike separated. We should either avoid such hypersurfaces or consider their dynamics only for a finite time until the Cauchy surface breaks.
    
    \item[b)] Regulators $\bq_\pm$ are chosen according to the IR regu\-larization described in Section~\ref{sec:EqualTimeSlice}. 
\end{itemize}

Regarding the finite regions $R$, the Cauchy surface breaking implies:
\begin{itemize}
    \item[c)] If the outermost inner point in the left wedge is $\bq_-^\up$, then the sign of its time coordinate is opposite to that of the IR regulator $\bq_-$. This fact can naively cause the Cauchy surface breaking. However, by sending the regulator to $i^0$, we effectively make the Cauchy surface breaking time infinite.
    
    \item[d)] If the outermost inner point in the left wedge is $\bq_-^\down$, then inevitably the corresponding Cauchy surface breaks, since the distance squared between $\bq_-^\down$ and $\bb_-^\up$ gets negative at some point (see Section~\ref{sec:Cauchy}).
\end{itemize}

If the conditions a) -- c) are met, then we get a Cauchy surface, on which complementarity and pure state condition are respected. If, instead of c), the option d) takes place, we obtain a hypersurface which gets timelike at some time moment and therefore, the entanglement entropy of Hawking radiation is well-defined only for a finite time.

%----------------------------%

\subsection{\label{sec:MS} Mirror-symmetric finite entangling region}

\begin{figure}[h]\centering
    \includegraphics[width=0.5\textwidth]{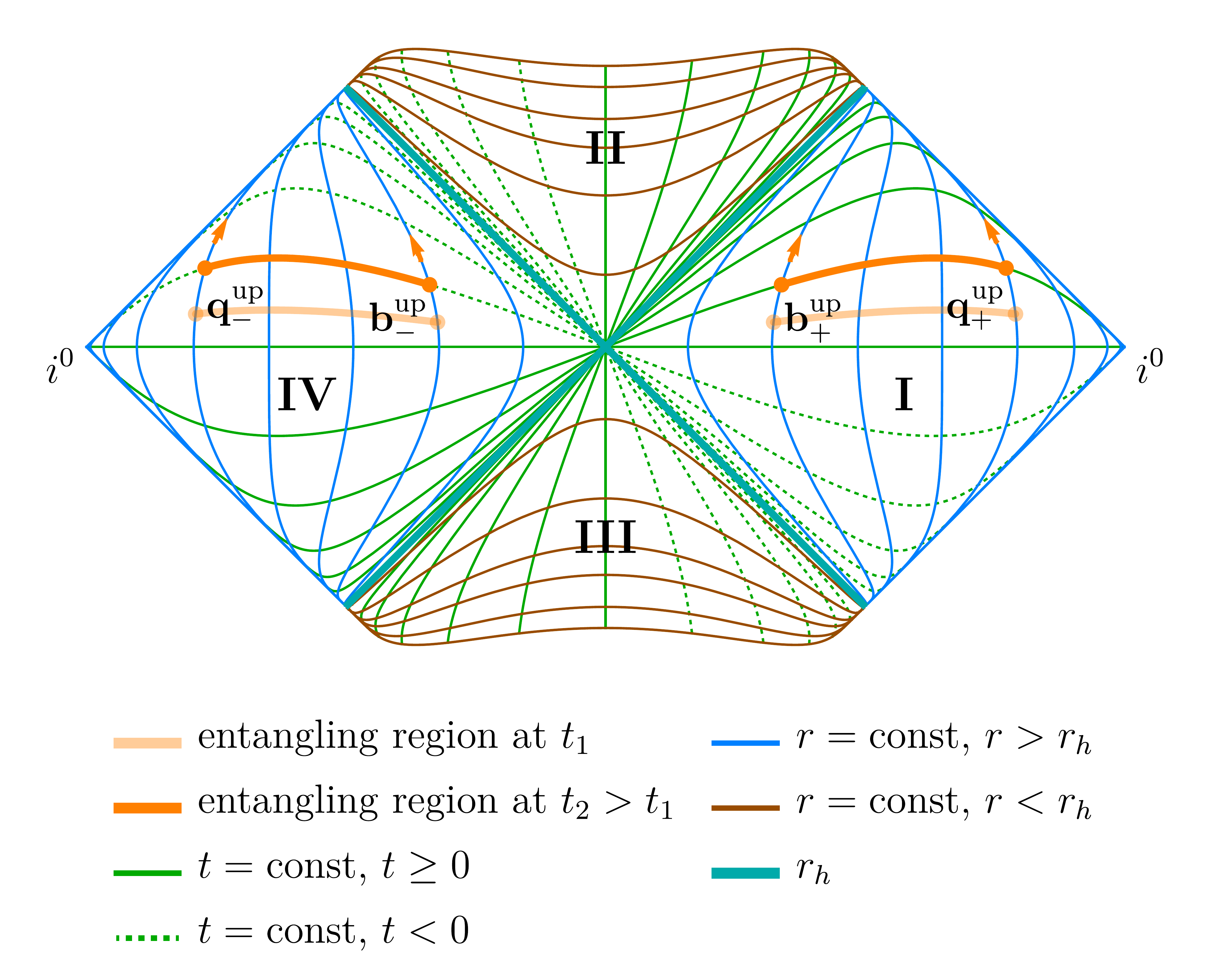}
    \caption{Penrose diagram for the eternal Schwarzschild black hole with a mirror-symmetric entangling region ${R_\MS \equiv [\bq_-^\up,\,\bb_-^\up] \cup [\bb_+^\up,\,\bq_+^\up]}$. Arrows indicate the direction of flow of points during time evolution.}
    \label{fig:FinRegNoIslUpUp}
\end{figure}

Let us consider the union of two finite intervals located in the right and left wedges, whose outer boundaries are mirror-symmetric\footnote{By mirror symmetry we mean the reflection about the vertical axis of symmetry of the Penrose diagram.} (see Fig.~\ref{fig:FinRegNoIslUpUp})
\be\nn
    R_\MS \equiv [\bq^{\up}_{-},\, \bb^{\up}_{-}] \cup [\bb^{\up}_{+},\, \bq^{\up}_{+}].
\ee
We call such a configuration the \textit{mirror-symmetric (MS) finite entangling region}.

The region $R_\MS$ has a straightforward interpretation. With respect to a static observer, one can imagine each part of $R_\MS$ as a domain between two concentric spheres with radii $b$ and $q > b$. Outgoing Hawking modes pass through this domain in a finite time and then escape to infinity.
\skipline

\begin{figure}[h]\centering
    \includegraphics[width=0.45\textwidth]{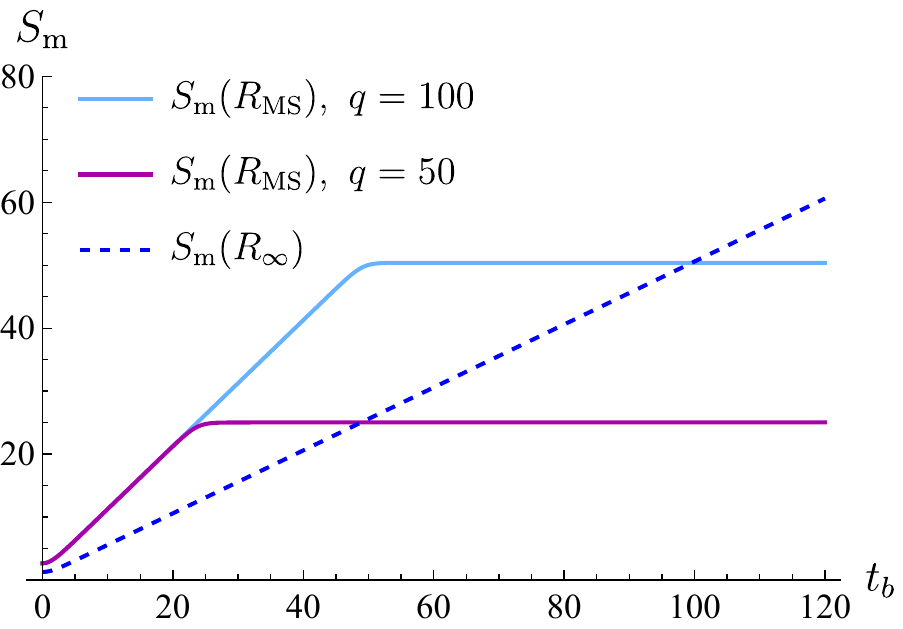}
    \caption{Entropy evolution of MS finite region $R_\MS$ with $q = 50$ (dark magenta), $q = 100$ (sky blue), and of the semi-infinite region $R_\infty$ (blue dashed). For all cases, we take $b = 5$, $r_h = 1$, $c = 3$, $\eps = 1$. The monotonic growth of $S_\m(R_\MS)$ is twice as fast compared to $S_\m(R_\infty)$.}
	\label{fig:Scompar}
\end{figure}

The entanglement entropy of Hawking radiation collected in the MS region is given by (see Appendix~\eqref{eq:app:MSnoISl})
\be
    \begin{aligned}
        S_\m(R_\MS) & = \frac{c}{6}\ln\left(\frac{16f(b)f(q)}{\kappa^4_h \eps^4}\right) + \frac{c}{3}\ln\cosh^2\kappa_h t_b + \\
        & + \frac{c}{3}\ln\left(\frac{\cosh\kappa_h(r_*(q) - r_*(b)) - 1}{\cosh\kappa_h(r_*(q) - r_*(b)) + \cosh 2\kappa_h t_b}\right).
    \end{aligned}
    \label{eq:qupnoisland}
\ee

If the region is large enough, i.e., $q \gg b$, then at intermediate times
\be\nn
    1 \ll \cosh 2\kappa_h t_b \ll \cosh \kappa_h (r_*(q) - r_*(b)),
\ee
the entanglement entropy of the radiation increases monotonically as
\be
    S_\m(R_\MS)\InterTimes \simeq \frac{2 c}{3}\,\kappa_h t_b,
    \label{eq:mongrowthUP}
\ee
which is twice as fast as for the semi-infinite region~\eqref{eq:late_time_without_island}. This behavior arises due to the fact that the distance between the points $\bq_-$ and $\bq_+$ is now time-dependent, since they lie on different timeslices. Also, provided that $t_b = t_q$, this time dependence is exactly the same as that of the distance between $\bb_-$ and~$\bb_+$, which doubles the coefficient.

At late times
\be\nn
    \cosh 2\kappa_h t_b \gg \cosh\kappa_h(r_*(q) - r_*(b)),
\ee
the entropy saturates at the value (see Appendix~\eqref{eq:app:MSnoISlLateTime})
\be
    \begin{aligned}
        S_\m(R_\MS)\LateTimes & \simeq \frac{c}{6}\ln\left(\frac{16f(b)f(q)}{\kappa_h^4\eps^4}\right) + \\ 
        & + \frac{c}{3}\ln\Big[\cosh\kappa_h(r_*(q) - r_*(b)) - 1\Big].
    \end{aligned}
    \label{eq:entropyfinitenoisland}
\ee
This can be interpreted as follows. As soon as the ``first''\footnote{Since the eternal black hole radiates permanently, by the ``first'' particle of radiation we mean the referent one emitted at $t = 0$.} particle of the black hole radiation reaches $r = b$, the entropy starts to increase, because more particles enter the domain between the spheres with radii $b$ and $q$. As this particle reaches $r = q$, the incoming and outgoing fluxes compensate each other and the entropy saturates.

The comparison of the entropies $S_\m(R_\infty)$~\eqref{eq:late_time_without_island} and $S_\m(R_\MS)$~\eqref{eq:qupnoisland} is demonstrated in Fig.~\ref{fig:Scompar}.

%----------------------------%

\subsubsection*{Islands of finite lifetime}

\begin{figure}[h]\centering
    \includegraphics[width=0.5\textwidth]{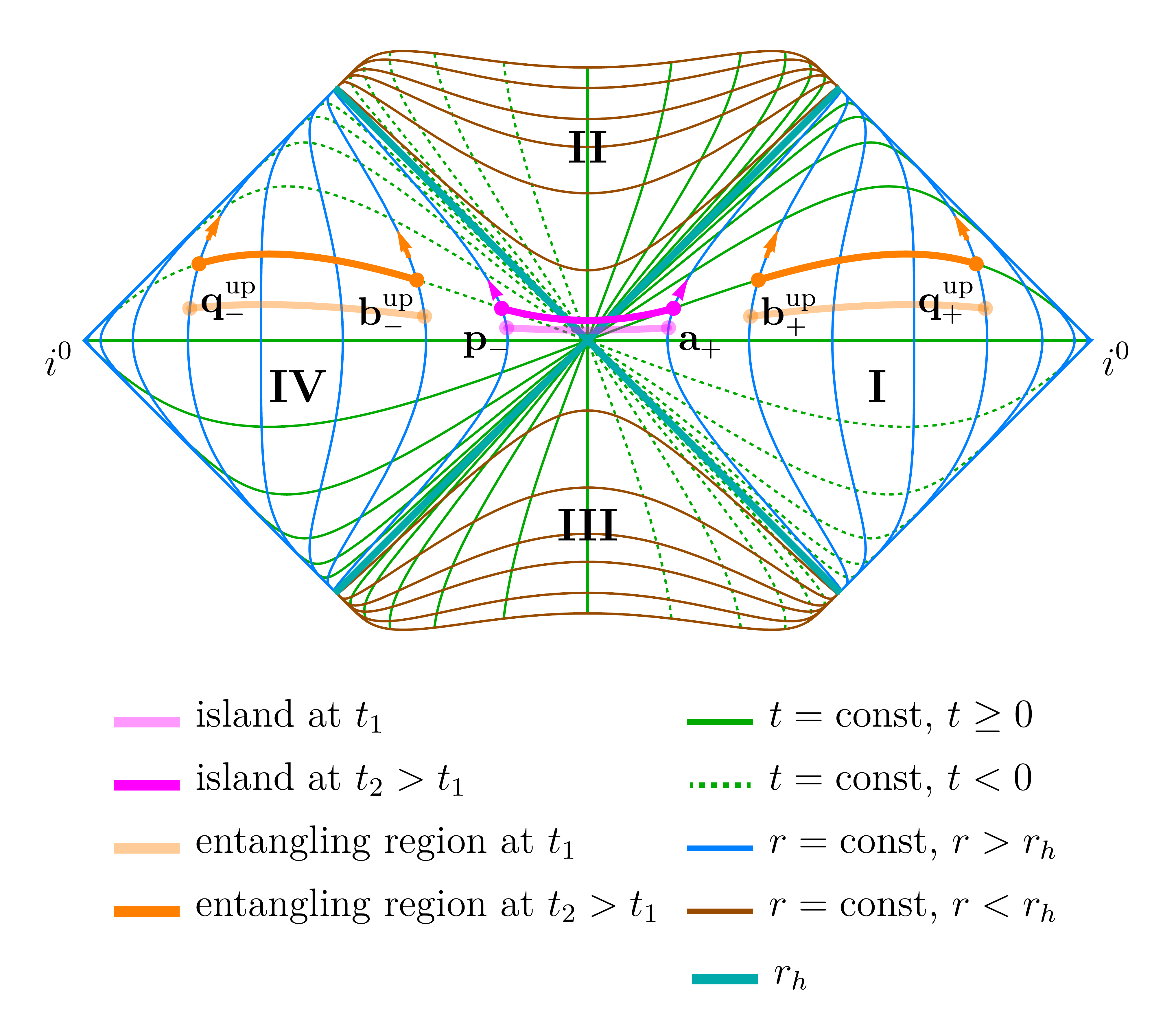}
    \caption{Penrose diagram for the eternal Schwarzschild black hole with a finite mirror-symmetric entangling region ${R_\MS \equiv [\bq_-^\up,\,\bb_-^\up] \cup [\bb_+^\up,\,\bq_+^\up]}$ (orange) and the island ${I = [\bp_-,\,\ba_+]}$ (magenta). Arrows indicate the direction of flow of points during time evolution.}
    \label{fig:FinRegWIslUpUp}
\end{figure}

Let us consider an arbitrary island configuration ${I = [\bp_-,\,\ba_+]}$, whose endpoints are parameterized~as (see Fig.~\ref{fig:FinRegWIslUpUp})
\be\nn
    \ba_+ = \left(a,\,t_a \right), \qquad \bp_- = \left(p,\,-t_p + \frac{i\pi}{\kappa_h}\right).
\ee

The full expression for $S_\gen[I, R_\MS]$ is given in Appendix~\ref{sec:appendix}, formula~\eqref{eq:app:MSwISl}. Since it is symmetric under permutations of the island coordinates: $a \leftrightarrow p$, $t_a \leftrightarrow t_p$, the extremization equations for~$a$ and $p$, as well as for~$t_a$ and $t_p$, are the same. This fact tells us that we should consider a mirror-symmetric ansatz for the island
\be\nn
    a = p, \qquad t_a = t_p.
\ee

For intermediate times
\be
    \begin{aligned}
        & \cosh\kappa_h (r_*(b) - r_*(a)) \ll \cosh\kappa_h(t_a + t_b) \ll \\
        & \ll \cosh\kappa_h (r_*(q) - r_*(a)),
    \end{aligned}
    \label{eq:conditiononislandUP}
\ee
there is an analytical solution to the extremization equations
\be\nn
    t_a = t_b, \qquad {0 < a - r_h \ll r_h}.
\ee
Using this, an approximate analytical expression for the entanglement entropy is given by (see Appendix~\eqref{eq:app:RmsAnFormEn} for details)
\be
    \begin{aligned}
        & S\extgen[I, R_\MS]\InterTimes \simeq S(R_\infty)\LateTimes + \frac{c}{3}\ln\cosh\kappa_h t_b - \\
        & - \frac{c}{3}\exp\Big[2 \kappa_h t_b - \kappa_h \left(r_*(q) - r_*(b)\right)\Big] + \frac{c}{6}\ln\frac{4f(q)}{\kappa^2_h \eps^2}.
    \end{aligned}
    \label{eq:RmsAnFormEn}
\ee
The first term on the RHS is the entropy for the semi-infinite region~\eqref{eq:late_time_with_island} studied in~\cite{HIM}. Other terms represent finite size effects. Interestingly, that the third term, being vanishingly small in the large-$q$ limit, affects the dynamics, although its influence is suppressed at intermediate times. The forth term gives the anomalous term~\eqref{eq:IR_anomaly} in the limit $q \to \infty$.

\begin{figure}[h]\centering
    \includegraphics[width=0.45\textwidth]{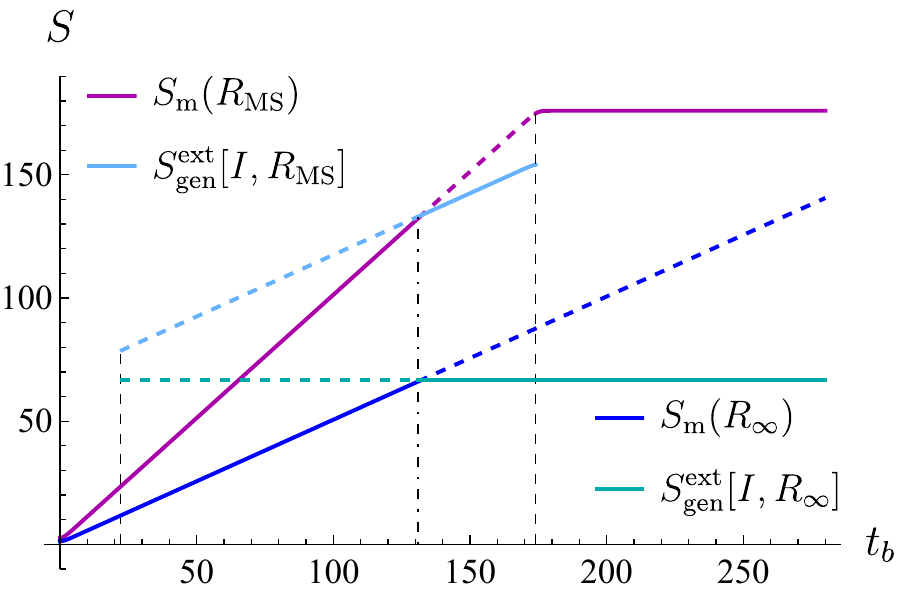}
	\caption{Entanglement entropy evolution of the finite entangling region $R_\MS$ with $q = 350$ (dark magenta, sky blue) and of the semi-infinite region $R_\infty$ (blue and cyan). For both cases, we take $b = 5$, $r_h = 1$, $c = 3$, $\GN = 0.1$, $\eps = 1$. The dominating (minimum) contribution is marked with solid lines. Non-dominating configurations are marked with dashed lines. The moments of the emergence and disappearance of the islands are marked with black dashed lines. After the disappearance of the island for the region $R_\MS$ (sky blue), there is an instantaneous transition to the entropy for the configuration without island (dark magenta).}
	\label{fig:largeqisland2}
\end{figure}

The growth of $S\extgen[I, R_\MS]$ is approximately linear and is the same as of the entropy~$S_\m(R_\infty)$~\eqref{eq:without_island}, unlike~$S_\m(R_\MS)$~\eqref{eq:mongrowthUP}. The comparison with numerical results is shown in Fig.~\ref{fig:largeqisland2}. 

At early times
\be\nn
    \cosh\kappa_h(t_a \pm t_b) \ll \cosh\kappa_h(r_*(b) - r_*(a)),
\ee
there is no real solution to the extremization problem for $a > r_h$, see~\eqref{eq:app:MSwIslEarlyTime}.

At late times
\be\nn
    \begin{aligned}
        \cosh\kappa_h(t_a + t_b) & \gg \cosh\kappa_h (r_*(q) - r_*(a)), \\
        \cosh 2\kappa_h t_b & \gg \cosh\kappa_h (r_*(q) - r_*(b)),
    \end{aligned}
\ee
along with $t_a \approx t_b$, the generalized entropy functional $S_\gen[I, R_\MS]$ grows monotonically with time $t_a$~\eqref{eq:app:MSwIslLargeTime}, hence there is no solution either.

\begin{figure}[h]\centering
    \includegraphics[width=0.45\textwidth]{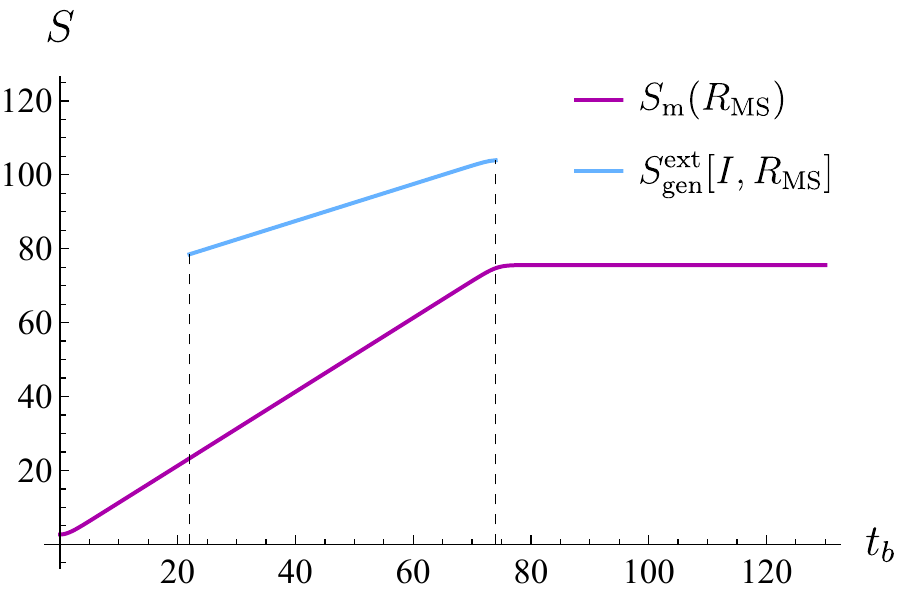}
	\caption{Entanglement entropy evolution of the region $R_\MS$ with $q = 150$ (dark magenta) and the generalized entanglement entropy functional $S\extgen[I, R_\MS]$ (sky blue). We take the parameters as $b = 5$, $r_h = 1$, $c = 3$, $\GN = 0.1$, $\eps = 1$. During the entire lifetime, the island does not dominate.}
	\label{fig:smallqisland2}
\end{figure}

\begin{figure}[h]\centering
   \includegraphics[width=0.45\textwidth]{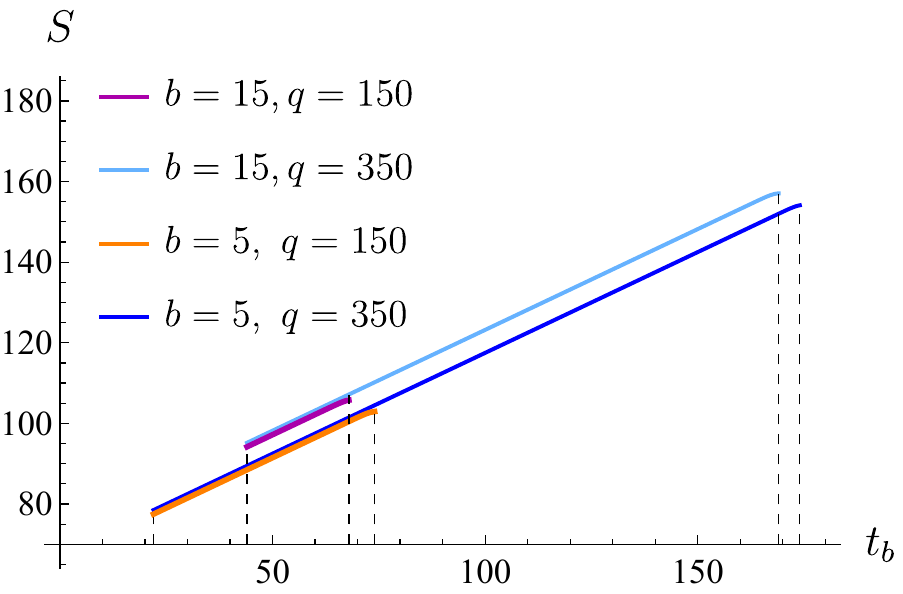}
	\caption{Time evolution of the generalized entanglement entropy $S\extgen[I, R_\MS]$ for different $b$ and $q$. The parameters are fixed as $r_h = 1$, $c = 3$, $\GN = 0.1$, $\eps = 1$. A decrease in~$q$ leads to a faster disappearance of the island. An increase in~$b$ leads to its later appearance.}
	\label{fig:diffislands2}
\end{figure}

We conclude that the island exists only for a \textit{finite time}, determined by the inequalities~\eqref{eq:conditiononislandUP} (see Fig.~\ref{fig:largeqisland2}). Starting from relatively small values of $q$, the island never dominates (see Fig.~\ref{fig:smallqisland2}). Reducing the size of the finite region~--- by increasing~$b$ or decreasing~$q$~--- leads to a decrease in the lifetime of the island (see Fig.~\ref{fig:diffislands2}). For sufficiently large $b$ or small $q$, the island does not appear at all, i.e., there is a region size threshold, at which the island ceases to exist.

There is also a \textit{discontinuity} in the entropy at the moment when the island disappears, see Fig.~\ref{fig:largeqisland2}. In Section~\ref{sec:IPFR}, we will show that this behavior causes the entanglement entropy to exceed the Bekenstein-Hawking entropy.

\begin{figure}[h]\centering
    \includegraphics[width=0.45\textwidth]{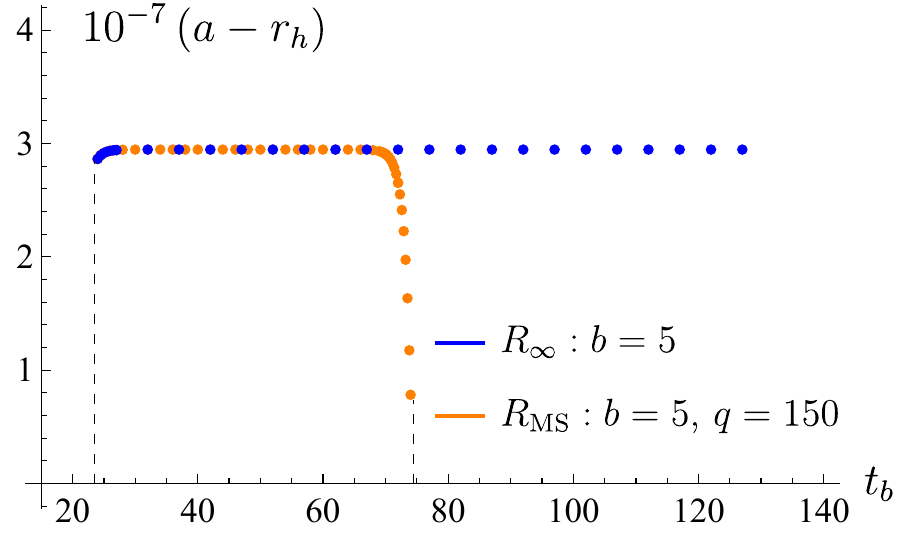}
    \includegraphics[width=0.45\textwidth]{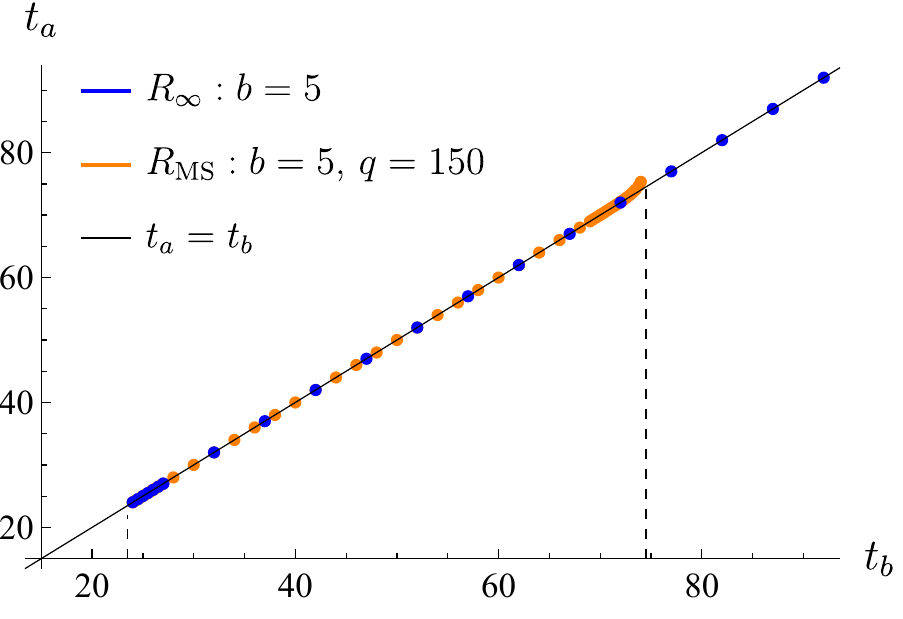}
	\caption{\textbf{Top:}~Evolution of the island radial coordinate~$a$ corresponding to the finite size entangling region $R_\MS$ (orange) with $q = 150$, and to the semi-infinite region $R_\infty$ (blue). \textbf{Bottom:}~Evolution of the island time coordinate $t_a$ for the same regions. Parameters are fixed as $b = 5$, $r_h = 1$, $c = 3$, $\GN = 0.1$, $\eps = 1$. Just before the island for the region $R_\MS$ disappears, its coordinates start to deviate from the coordinates of the island for $R_\infty$.}
	\label{fig:islandevol2}
\end{figure}

Numerical analysis reveals several additional features. From Fig.~\ref{fig:largeqisland2} and~\ref{fig:smallqisland2}, we note that the entropy for the configuration without islands reaches saturation at the moment that approximately coincides with the disappearance of the island. Also, the parameters of the island configuration~$a$ and $t_a$ for $R_\MS$ coincides with that for $R_\infty$, up to a short period of time before the disappearance of the island (see Fig.~\ref{fig:islandevol2}). The island gets smaller, but remains mirror-symmetric throughout the entire lifetime.

%----------------------------%

\subsection{Asymmetric finite entangling region}

\subsubsection*{\label{sec:Cauchy}Cauchy surface breaking}

\begin{figure}[h]\centering
    \includegraphics[width=0.5\textwidth]{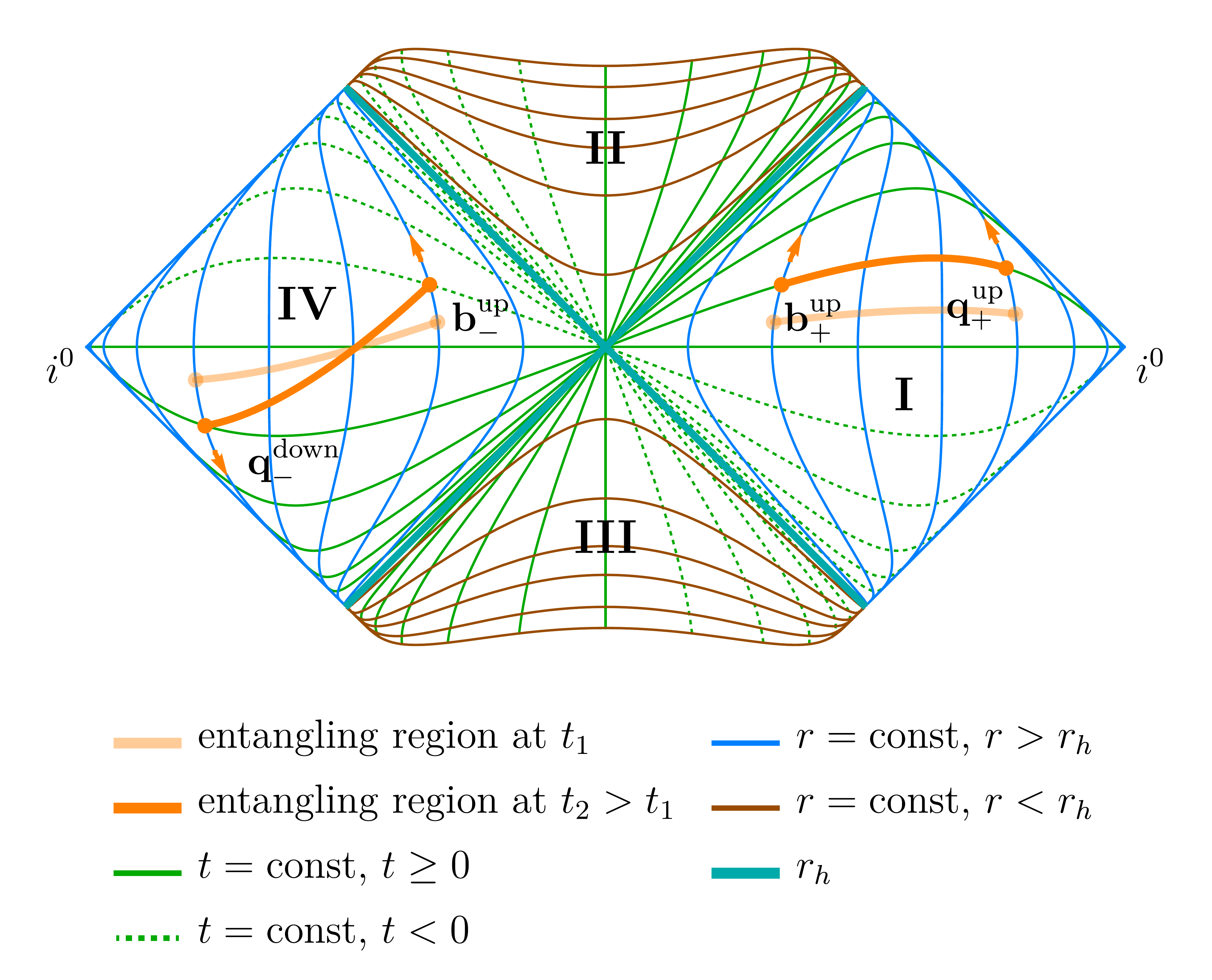}
    \caption{Penrose diagram for the eternal Schwarzschild black hole with a finite asymmetric entangling region ${R_\AS \equiv [\bq_-^\down,\,\bb_-^\up] \cup [\bb_+^\up,\,\bq_+^\up]}$. Arrows indicate the direction of flow of points during time evolution.}
    \label{fig:FinRegNoIslUpDown}
\end{figure}

Now let us consider the following union of two intervals, which we call the \textit{asymmetric (AS) finite entangling region} (see Fig.~\ref{fig:FinRegNoIslUpDown})
\be\nn
    R_\AS \equiv [\bq^\down_{-},\,\bb^\up_{-}] \cup [\bb^\up_{+},\,\bq^\up_{+}].
\ee
Its name comes from the fact that this region is not in any sense symmetrical in the Penrose diagram.

In this setup, there is an \textit{upper bound} on time $t_b$ (see Section~\ref{sec:pick-Cauchy}): during time evolution, the point $\bq_-^\down$ moves in time along the flow of the Killing vector $\partial_t^-$, while the point $\bb_-^\up$ moves in the opposite direction, such that the interval between them eventually becomes timelike (see Fig.~\ref{fig:FinRegNoIslUpDown}). Indeed, consider the distance squared between~$\bq_-^\down$ and~$\bb_-^\up$
\be\nn
    d^{\,2}\left(\bq^\down_-,\,\bb^\up_-\right) \propto \cosh\kappa_h(r_*(q) - r_*(b)) - \cosh 2 \kappa_h t_b.
\ee
This expression gets negative at the moment
\be
    \tbreak(b, q) = \frac{r_*(q) - r_*(b)}{2}.
    \label{eq:breaktime}
\ee

\begin{figure}[h]\centering
    \includegraphics[width=0.45\textwidth]{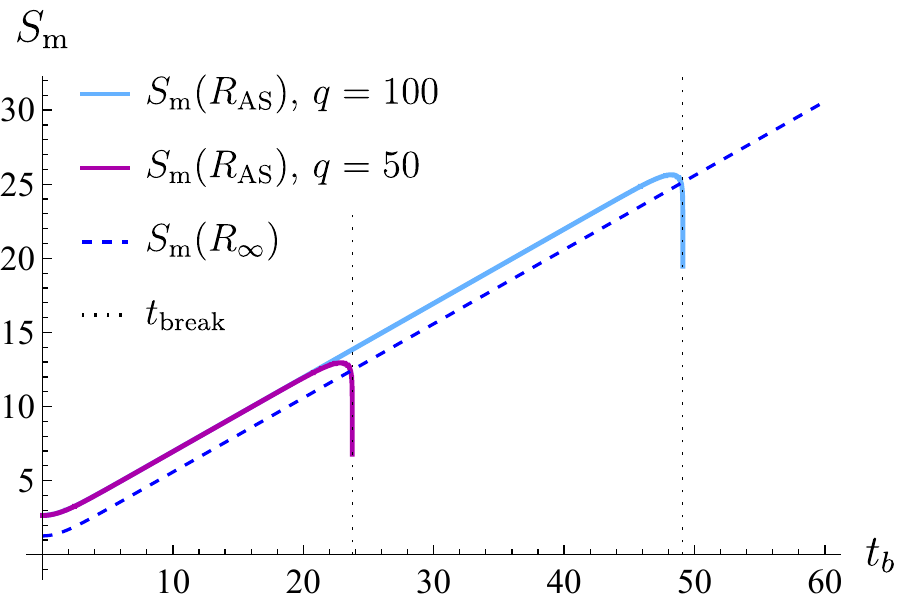}
    \caption{Entropy evolution for the finite entangling region $R_\AS$ with $q = 50$ (dark magenta), $q = 100$ (sky blue) and for the semi-infinite one $R_\infty$ (blue). For all cases, we take ${b = 5}$, ${r_h = 1}$, ${c = 3}$, ${\eps = 1}$. The entropy for $R_\AS$  has the same slope as that for~$R_\infty$, but gets singular at $\tbreak$~\eqref{eq:breaktime} (dotted lines).}
	\label{fig:qdownoisland}
\end{figure}

Hence, for $t_b > \tbreak$, the problem becomes ill-defined, since there is no longer a Cauchy surface (i.e. a spacelike hypersurface) to define a pure quantum state. A larger size of the finite region~$R_\AS$ leads to a larger~$\tbreak$. In the limit $q \to \infty$, this time also gets infinite: $\tbreak \to \infty$, and the Cauchy surface breaking never happens.

For intermediate times $r_h \ll t_b \ll \tbreak$, the entanglement entropy for~$R_\AS$ grows linearly (refer to Appendix~\eqref{eq:app:ASnoISl} for the full expression)
\be
    S_\m(R_\AS) \simeq \frac{c}{3}\,\kappa_h t_b.
    \label{eq:mongrowth}
\ee

Fig.~\ref{fig:qdownoisland} shows that the entanglement entropy for $R_\AS$ almost coincides with that for the semi-infinite region $R_\infty$, without strong dependence on~$q$. Just before the breaking time $\tbreak$, the entropy abruptly decreases and hits the singularity, after which it is not well-defined.

%----------------------------%

\subsubsection*{Non-symmetric islands}

\begin{figure}[h]\centering
    \includegraphics[width=0.5\textwidth]{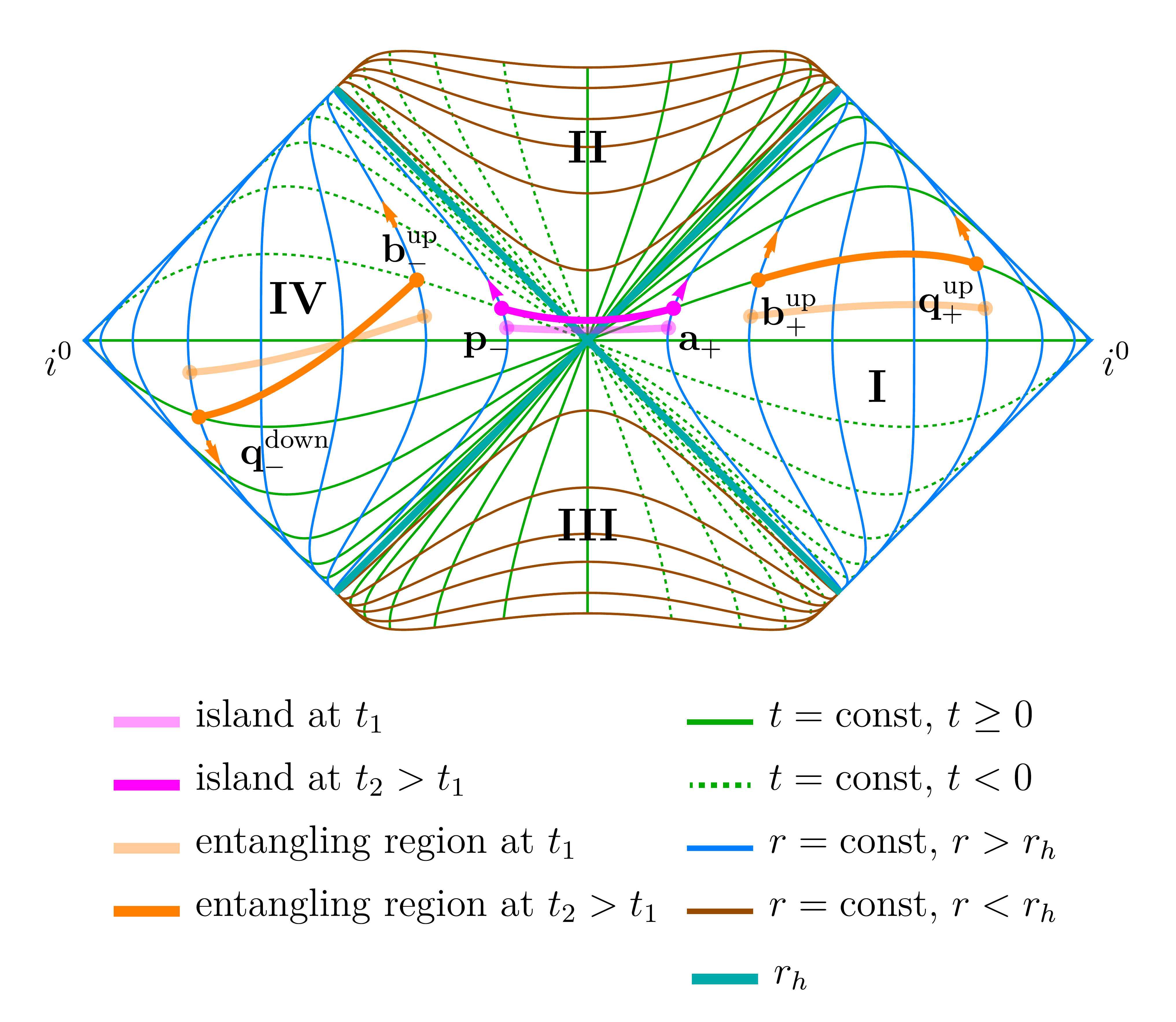}
    \caption{Penrose diagram for the eternal Schwarzschild black hole with a finite asymmetric entangling region ${R_\AS \equiv [\bq_-^\down,\,\bb_-^\up] \cup [\bb_+^\up,\,\bq_+^\up]}$ (orange) and the island ${I = [\bp_-,\,\ba_+]}$ (magenta). Arrows indicate the direction of flow of points during time evolution.}
    \label{fig:FinRegWIslUpDown}
\end{figure}

Let us consider the island ansatz $I = [\bp_-,\,\ba_+]$ for the AS entangling region (see Fig.~\ref{fig:FinRegWIslUpDown}). For relatively early times,~i.e.
\be\nn
    \begin{aligned}
        & \cosh\kappa_h(t_p + t_b) \ll \cosh\kappa_h(r_*(q) - r_*(p)), \\
        & \cosh\kappa_h(t_a - t_b) \ll \cosh\kappa_h(r_*(q) - r_*(a)),
    \end{aligned}
\ee
all the terms in the formula for the generalized entropy functional $S_\gen[I, R_\AS]$, that are non-symmetric with respect to permutations of the island coordinates: ${a \leftrightarrow p}$, ${t_a \leftrightarrow t_p}$, get suppressed (see Appendix~\eqref{eq:app:ASwISl}). Hence, for early times, it is reasonable to take a mirror-symmetric ansatz: ${a \simeq p}$, ${t_a \simeq t_p}$.

Then, for intermediate times
\be
    \begin{aligned}
        & \cosh\kappa_h(r_*(a) - r_*(b)) \ll \cosh\kappa_h(t_a + t_b) \ll \\
        & \ll \cosh\kappa_h(r_*(q) - r_*(b)),
    \end{aligned}
    \label{eq:AS_inter_times}
\ee
the extremization gives the same solution as in~\cite{HIM}
\be\nn
    t_a = t_b, \qquad 0 < a - r_h \ll r_h.
\ee
The approximate analytical expression for the generalized entropy is given by~\eqref{eq:app:ASwISlInterTime}
\be
    \begin{aligned}
        & S\extgen[I, R_\AS]\InterTimes \simeq S(R_\infty)\LateTimes - \\
        & - \frac{c}{3}\exp\Big[2 \kappa_h t_b - \kappa_h \left(r_*(q) - r_*(b)\right)\Big] + \frac{c}{6}\ln\frac{4f(q)}{\kappa^2_h \eps^2}.
    \end{aligned}
    \label{eq:RasAnFormEn}
\ee
The limit $q \to \infty$ at fixed $t_b$ of this expression recovers the semi-infinite case~\eqref{eq:late_time_with_island}, except for the anomalous term~\eqref{eq:IR_anomaly}, which is to be subtracted.

\begin{figure}[h]\centering
    \includegraphics[width=0.45\textwidth]{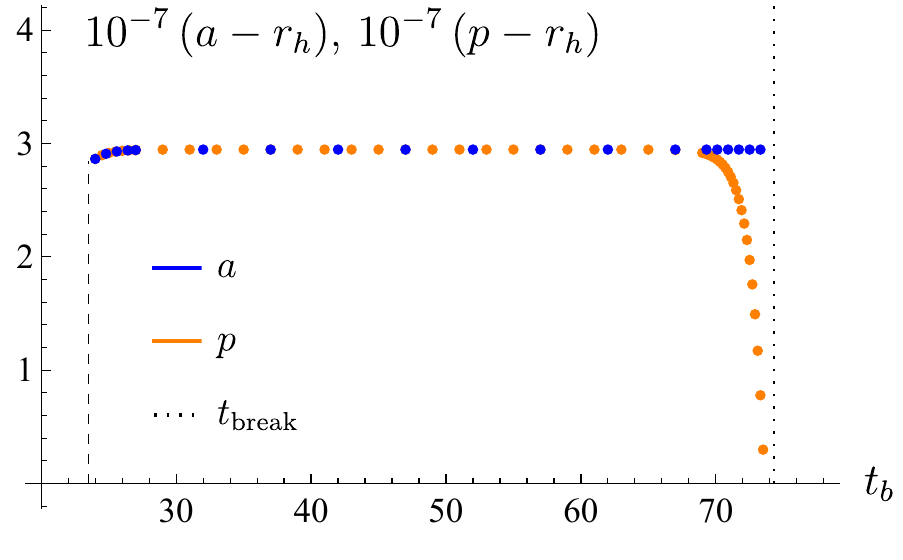}
    \includegraphics[width=0.45\textwidth]{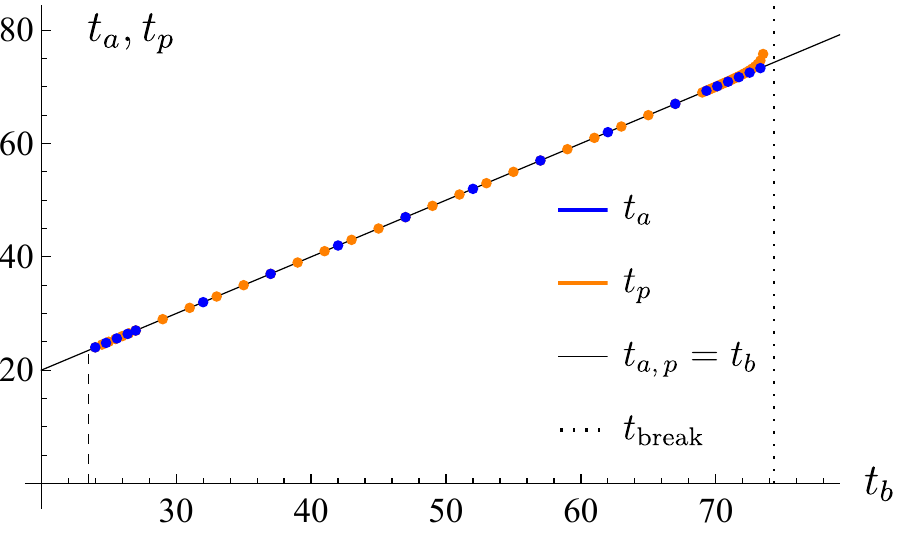}
	\caption{\textbf{Top:}~Evolution of the island radial coordinates~$a$ (blue) and $p$ (orange) corresponding to the finite size entangling region $R_\AS$ with $b = 5$, $q = 150$. \textbf{Bottom:}~Evolution of the island time coordinates~$t_a$ (blue) and $t_p$ (orange) corresponding to the same region. Parameters are fixed as $r_h = 1$, $c = 3$, $\GN = 0.1$, $\eps = 1$. Near the breaking time $\tbreak$~\eqref{eq:breaktime}, there is spatial $a \neq p$ and time $t_a \neq t_p$ asymmetries.}
	\label{fig:islandevol}
\end{figure}

At late times
\be\nn
    \begin{aligned}
        & \cosh\kappa_h(t_p + t_b) \gg \cosh\kappa_h(r_*(q) - r_*(p)), \\
        & \cosh\kappa_h(t_a - t_b) \gg \cosh\kappa_h(r_*(q) - r_*(a)),
    \end{aligned}
\ee
the contribution of non-symmetric terms in $S_\gen[I, R_\AS]$ becomes significant, which leads to the fact that the solution to the extremization equations becomes non-symmetric. This logic is verified by numerical calculations, see Fig.~\ref{fig:islandevol}.

\begin{figure}[h]\centering
    \includegraphics[width=0.45\textwidth]{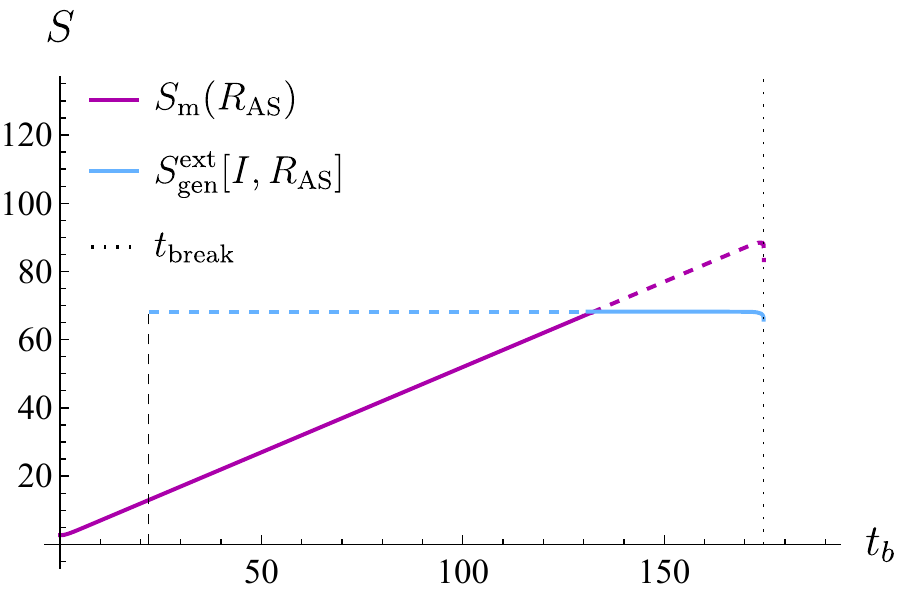}
    \caption{Entanglement entropy evolution of the region $R_\AS$ with $b = 5$ and $q = 350$ (dark magenta), and of the same region with the island (sky blue). The parameters are taken as ${r_h = 1}$, ${c = 3}$, ${\GN = 0.1}$ and ${\eps = 1}$. Note the abrupt decrease near $\tbreak$~\eqref{eq:breaktime}, which is caused by the subsequent Cauchy surface breaking.}
	\label{fig:largeqisland}
\end{figure}

\begin{figure}[h]\centering
    \includegraphics[width=0.45\textwidth]{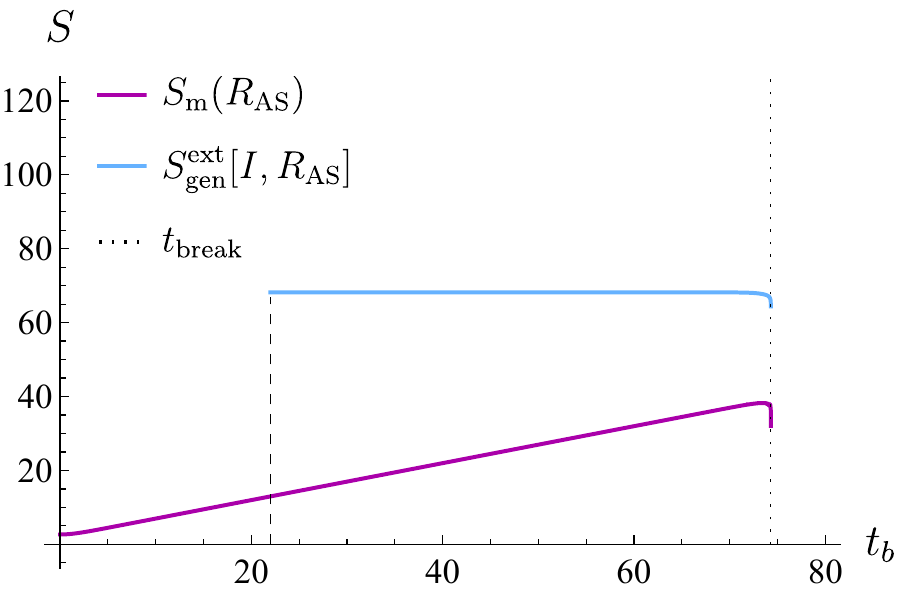}
	\caption{Entanglement entropy evolution of the region $R_\AS$ with $b = 5$ and $q = 150$ (dark magenta), and of the same region with the island (sky blue). The parameters are taken as $r_h = 1$, $c = 3$, $\GN = 0.1$ and $\eps = 1$. The extremization procedure never leads us the entropy with an island over that for the configuration without island.}
	\label{fig:smallqisland}
\end{figure}

Numerical results provide us with more details. For large finite entangling regions $r_*(q) \gg r_*(b)$ (see Fig.~\ref{fig:largeqisland}), the dynamics of the entropy is qualitatively the same as for $R_\infty$, except that $S(R_\AS)$ slightly decreases just before the Cauchy surface breaking. For smaller regions, the island contribution is never dominant (see Fig.~\ref{fig:smallqisland}).

The Cauchy surface breaking, which bounds the lifetime of the configuration containing AS finite region, constrains the lifetime of the island as well. It is not the case for sufficiently large regions, such that $\cosh\kappa_h(r_*(a) - r_*(b)) \ll \cosh\kappa_h \tbreak$, because in this case, there is no solution under the condition~\eqref{eq:AS_inter_times}. However, for smaller regions, the appearance of the island may not occur before $\tbreak$.

\begin{figure}[h]\centering
    \includegraphics[width=0.45\textwidth]{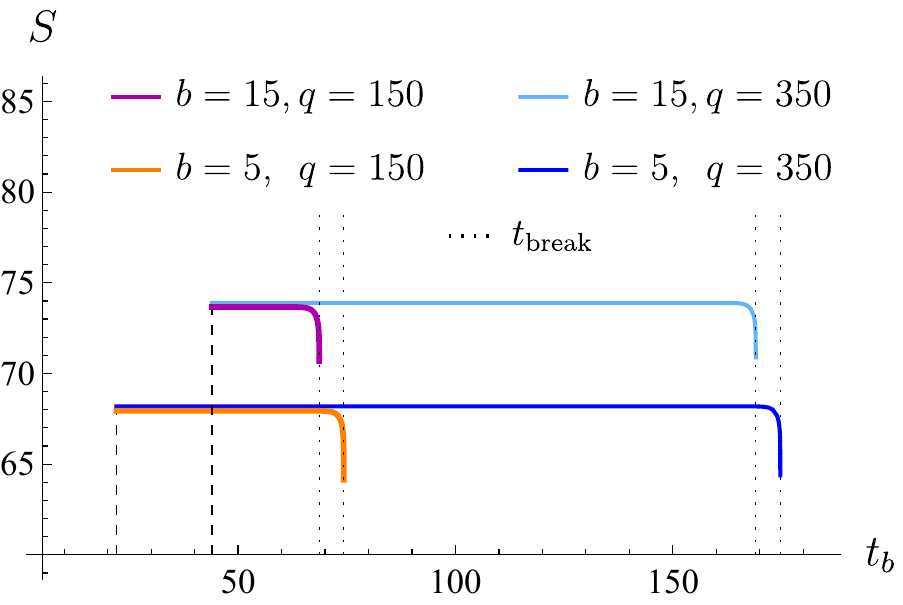}
    \caption{Time evolution of the generalized entanglement entropy $S\extgen[I, R_\AS]$ for different $b$ and $q$. Parameters are fixed as $r_h = 1$, $c = 3$, $\GN = 0.1$, $\eps = 1$. The lifetime of the island configuration increases due to the change in the time of the Cauchy surface breaking $\tbreak$, as well as due to the dependence of the time of the island appearance on~$b$.}
	\label{fig:diffislands}
\end{figure}

The lifetime of the island shortens as the size of the entangling region $R_\AS$ gets smaller (as $q$ decreases or $b$ increases, see Fig.~\ref{fig:diffislands}).

%%%%%%%%%%%%%%%%%%%%%%%%%%%%%%%%%%%%%%%%%%%%%%%%

\section{\label{sec:IPFR}Information loss paradox for finite entangling regions}

A Cauchy surface~$\Sigma$ in the eternal Schwarzschild black hole might be divided into the semi-infinite entangling region~$R_\infty$, the region associated with the black hole~$BH$, and the domain in-between. If the latter is negligible, we have
\be
    \Sigma = BH \cup R_\infty.
\ee

Given that the state is pure, the entanglement entropy obeys the complementarity property
\be
    S(R_\infty) = S(BH).
    \label{eq:compropertyforIP}
\ee

The fine-grained entropy~$S(BH)$ is to be bounded from above by the coarse-grained entropy~$S_\text{thermo}(BH)$
\be
    S(BH) \leq S_\text{thermo}(BH).
    \label{eq:coarsse-grained-for-BH}
\ee
The latter is twice the Bekenstein-Hawking entropy \cite{Almheiri:2019yqk, HIM}
\be
    S_\text{thermo}(BH) = 2 S_\BH = \frac{2\pi r^2_h}{\GN}.
    \label{eq:coarseforeternal}
\ee
As a result, the upper bound on the entanglement entropy of Hawking radiation is expected to be~\cite{Almheiri:2020cfm}
\be
    S(R_\infty) \leq 2 S_\BH.
    \label{eq:IPforInfinite}
\ee
In fact, this limit is violated due to the unstoppable growth of the entanglement entropy. This might be seen as a version of the information loss paradox~\cite{HIM, Almheiri:2019yqk}.

The evolution of the entropy changes in the presence of entanglement islands. When the island starts to dominate, the entanglement entropy is given by~\eqref{eq:late_time_with_island} and consists of two terms, one of which is the Bekenstein-Hawking entropy, while the other denotes additional corrections
\be
    S(R_\infty)\LateTimes \simeq 2 S_\BH + S_\text{corr}.
\ee
These corrections are time-independent and small compared to the area term $S_\BH$ under the ``black hole classicality'' condition~\cite{HIM}
\be
    \frac{r^2_h}{\GN} \gg c.
    \label{eq:BH-classicality}
\ee
We say that the information paradox in two-sided Schwarzschild black hole does not arise if either the bound~\eqref{eq:IPforInfinite} is respected or violated only by terms suppressed under~\eqref{eq:BH-classicality}. Formally,
\be
    S(R_\infty) \leq 2 S_\BH + S_\text{corr}, \qquad \frac{S_\text{corr}}{S_\BH} \ll 1.
\ee
\skipline

\begin{figure}[h]\centering
    \includegraphics[width=0.5\textwidth]{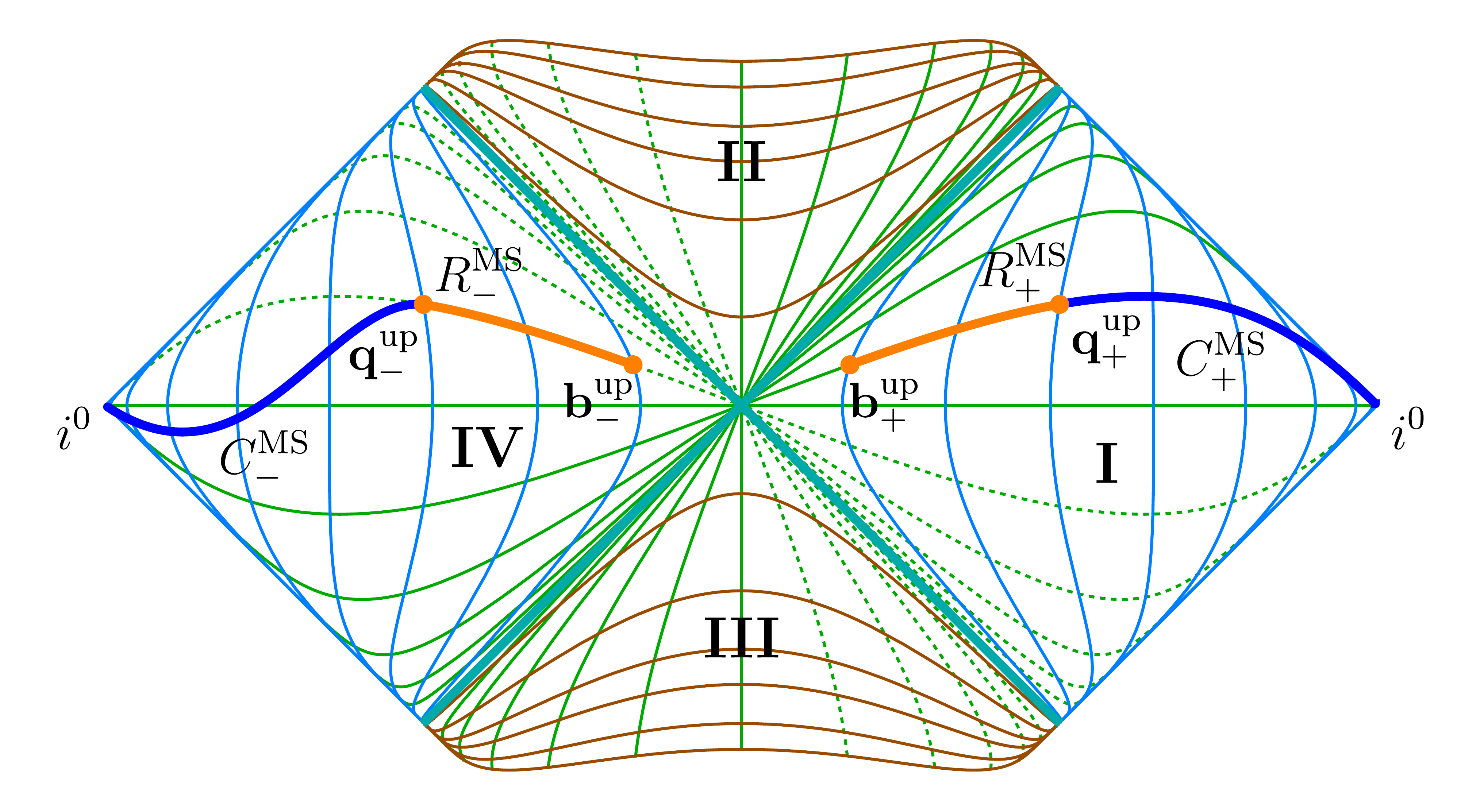}
    \includegraphics[width=0.5\textwidth]{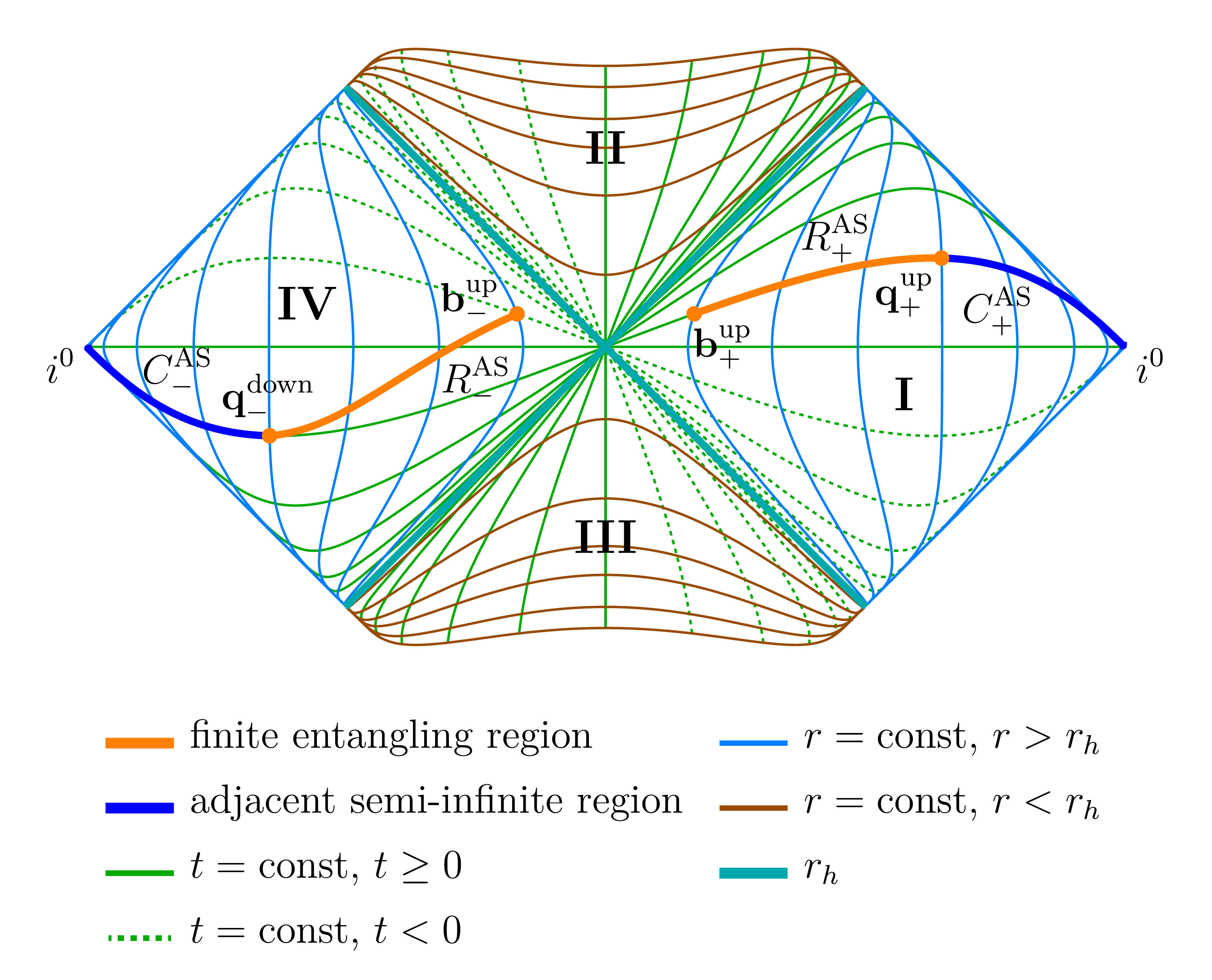}
	\caption{\textbf{Top:}~Penrose diagram for the eternal Schwarzschild black hole with semi-infinite entangling region $R_\infty$ partitioned into finite MS subregion, ${R_\MS \equiv R^\MS_- \cup R^\MS_+}$ (orange), and its adjacent semi-infinite region ${C_\MS \equiv C^\MS_- \cup C^\MS_+}$ (blue). \textbf{Bottom:}~The same diagram with semi-infinite entangling region $R_\infty$ partitioned into finite AS subregion, ${R_\AS \equiv R^\AS_- \cup R^\AS_+}$ (orange), and its adjacent semi-infinite region ${C_\AS \equiv C^\AS_- \cup C^\AS_+}$ (blue).}
	\label{fig:CAS-CMS}
\end{figure}

We can also divide a Cauchy surface into the region associated with the black hole~$BH$, a finite entangling region~$R$, a finite domain in-between and an adjacent semi-infinite region~$C$, which extends to spacelike infinities $i^0$ (see Fig.~\ref{fig:CAS-CMS})
\be
    \Sigma = BH \cup R \cup C.
    \label{eq:tripartition}
\ee

The strong subadditivity of entanglement entropy~\cite{Nishioka:2018khk} for a tripartition like~\eqref{eq:tripartition} gives the inequality
\be
    S(BH \cup R \cup C) + S(R) \leq S(BH \cup R) + S(R \cup C).
\ee
Then, using the complementarity property
\be
    \begin{aligned}
        & S(BH) = S(R \cup C), \\
        & S(R) = S(BH \cup C), \\
        & S(C) = S(BH \cup R),
    \end{aligned}
    \label{eq:equalityoftripart}
\ee
and pure state condition for the total state, $S(\Sigma) = 0$, we derive the upper bound on the entanglement entropy for a finite region~$R$ \textit{(the strong bound)}
\be
    S(R) \leq 2 S_\BH + S(C).
    \label{eq:IPFinite}
\ee
We interpret its violation as the information paradox for finite entangling regions. The island prescription leads to softening of this constraint \textit{(the soft bound)}
\be
    S(R) \leq 2 S_\BH + S(C) + S_\text{corr}.
    \label{eq:IPFinite_soft}
\ee

%-------------------------------------%

\subsection{MS finite entangling region}

\subsubsection*{Do islands for \texorpdfstring{$C_\MS$}{C-MS} influence the bound?}

As $C_\MS$ we referred to a semi-infinite outer entangling region, which is adjacent to $R_\MS$. It can be defined as the limit of the following finite entangling region
\be\nn
    C_\MS = \lim\limits_{w \to \infty}[\bw_-^\down,\,\bq_-^\up] \cup [\bq_+^\up,\,\bw_+^\up].
\ee
The points $\bw_-^\text{down}$ and $\bw_+^\text{up}$ are IR regulators. Essentially, it is the same region as considered in~\cite{HIM}. Hence, we already know the features of its dynamics: the entropy for $C_\MS$ grows at early times (see~\eqref{eq:without_island})
\be
    S(C_\MS)\EarlyTimes = \frac{c}{6}\ln\left(\frac{4f(q)\cosh^2\kappa_h t_q}{\kappa^2_h\eps^2}\right),
    \label{eq:cms_early_times}
\ee
while at late times, it saturates due to the formation of the island $I_{C}$ for this region (see~\eqref{eq:late_time_with_island})
\be
    S(C_\MS)\LateTimes \simeq \frac{2\pi r_h^2}{\GN} + \frac{c}{6}\ln\frac{f(q)}{\kappa^4_h\eps^4} + \frac{c}{6}\left(2\kappa_h r_*(q) - 1\right).
    \label{eq:cms_late_times}
\ee

Since the endpoints $\bq_\pm^\up$ are adjacent for $R_\MS$ and $C_\MS$, their shifts influence oppositely the formation of the islands $I$ for $R_\MS$ and $I_C$ for $C_\MS$. Indeed, there are the island solutions, when the following conditions are satisfied
\be\nn
    \begin{aligned}
        \text{$R_\MS$: } & \cosh\kappa_h(r_*(b) - r_*(a^R)) \ll \cosh\kappa_h(t_a^R + t_b) \ll \\
        & \ll \cosh\kappa_h(r_*(q) - r_*(a^R)), \\
        \text{$C_\MS$: } & \cosh\kappa_h(t_a^C + t_b) \gg \cosh\kappa_h (r_*(q) - r_*(a^C)) > \\
        & > \cosh\kappa_h (r_*(q) - r_*(b)).
    \end{aligned}
\ee
We see that these conditions cannot hold together because $a^R \simeq a^C$, therefore, the corresponding islands $I$ and $I_C$ do not exist simultaneously.

Actually, even when the island $I_C$ is formed, the bound on the entropy for $R_\MS$, imposed by $S(C_\MS)\InTextLateTimes$, cannot be violated, because $S(C_\MS)\InTextLateTimes$~\eqref{eq:cms_late_times} and $S_\m(R_\MS)\InTextLateTimes$~\eqref{eq:entropyfinitenoisland} have the same dependence on $q$, thus never intersect. What we are left with to check is whether the strong bound, related to $C_\MS$ without the island $I_C$, holds
\be
    S(R_\MS) \overset{\Huge\mathbf{?}}{\leq} S_\text{bound} = \frac{2\pi r_h^2}{\GN} + \frac{c}{6}\ln\left(\frac{4f(q)\cosh^2\kappa_h t_q}{\kappa^2_h\eps^2}\right).
    \label{eq:upIP1}
\ee

%-------------------------------------%

\subsubsection*{Without islands for \texorpdfstring{$R_\MS$}{R-MS}}

\begin{figure}[h]\centering
    \includegraphics[width=0.45\textwidth]{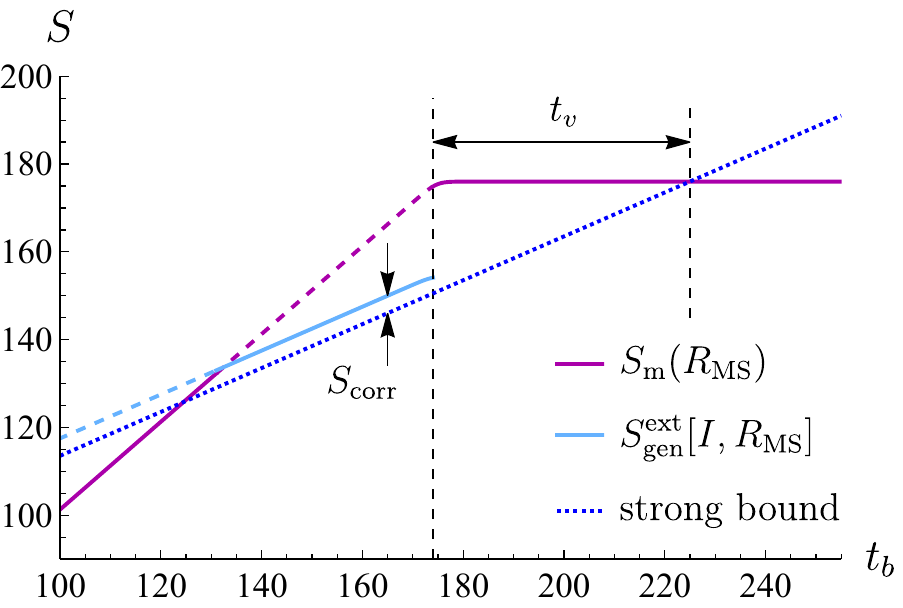}
	\caption{The early-time evolution of the matter entropy $S_\m(R_\MS)$ (dark magenta), the entropy for the configuration with the island $S\extgen[I, R_\MS]$ (sky blue) and the strong bound~\eqref{eq:upIP1} (dotted blue) for a finite size entangling region $R_\MS$ with $b = 5$ and $q = 350$. We take ${r_h = 1}$, ${c = 3}$, ${\GN = 0.1}$, ${\eps = 1}$. After the disappearance of the island, there is a discontinuous transition to the entropy of matter $S_\m(R_\MS)$ that is larger than the upper bound during some finite time of violation~$t_v$. As $q$ increases, the maximum difference between $S_\m(R_\MS)$ and the strong bound increases, while $S_\text{corr}$ does not change significantly.}
	\label{fig:IPq350}
\end{figure}

Since the entropy of matter for MS region~\eqref{eq:mongrowthUP} grows twice as fast as that for the semi-infinite region $R_\infty$, and the strong bound~\eqref{eq:upIP1} is imposed by the semi-infinite region of the same type as $R_\infty$, the bound might be eventually violated. However, the entropy saturates at some finite value (see Section~\ref{sec:MS}), while the bound would keep growing linearly. Therefore, if the bound is violated, then only for a finite time (see dark magenta curve in Fig.~\ref{fig:IPq350}).

%-------------------------------------%

\subsubsection*{Island dominating for finite time}

When the island $I$ dominates, $S\extgen[I, R_\MS]$ is larger than the strong bound by $S_\text{corr}$, which does not change significantly if we vary $q$. Indeed, their difference~$S_\text{corr}$ is
\be
    \begin{aligned}
        S_\text{corr} & = S\extgen[I, R_\MS] - S_\text{bound} \simeq \\
        & \simeq \frac{c}{6}\ln\frac{f(b)}{\kappa^4_h\eps^4} + \frac{c}{3}\kappa_h r_*(b) - \frac{c}{6} - \\
        & - \frac{c}{3}\exp\Big[2 \kappa_h t_b - \kappa_h \left(r_*(q) - r_*(b)\right)\Big],
    \end{aligned}
    \label{eq:Scorr_MS}
\ee
with a strongly suppressed dependence on $t_b$ and $q$.

After the disappearance of the island, there is a discontinuous transition to the constant entropy~\eqref{eq:entropyfinitenoisland}. Its value depends on $q$ and is significantly larger than the strong bound (see Fig.~\ref{fig:IPq350}), because during the island domination, the matter entropy $S_\m(R_\MS)$ grows twice as fast~\eqref{eq:mongrowthUP} as the bound. After that, the difference decreases. The larger the value of $q$ is~--- the longer the constraint is violated. Thus, we say that the island prescription for $R_\MS$ does \textit{not} solve the information paradox completely.

%-------------------------------------%

\subsubsection*{Never dominant island}

For relatively small sizes of the region $R_\MS$, the entropy of matter always dominates (see Section~\ref{sec:MS}) and does not exceed the entropy with an island
\be
    S(R_\MS) = S_\m(R_\MS) < S\extgen[I, R_\MS] = S_\text{bound} + S_\text{corr},
\ee
hence, the soft bound is not violated.

%-------------------------------------%

\subsection{AS finite entangling region}

The semi-infinite adjacent region for AS region turns out to be neither of MS type nor AS. It can be defined as the following limit
\be\nn
    C_\AS = \lim\limits_{w \to \infty}[\bw_-^\down,\,\bq_-^\down] \cup [\bq_+^\up,\,\bw_+^\up].
\ee
The entropy of matter for this region is (see Appendix~\eqref{eq:app:CAS})
\be
    S_\m(C_\AS) = \frac{c}{6}\ln\frac{4f(q)}{\kappa^2_h \eps^2},
    \label{eq:noislandcregiondown}
\ee
which is time-independent, because all points of $C_\AS$ lie on the same timeslice. This entropy is of order of ${c \ll r_h^2/\GN}$, and hence, it is subdominant compared to $S_\BH$. This means that there is no information paradox for the region $C_\AS$ even without islands.

As for the region $R_\AS$, the strong bound~\eqref{eq:IPFinite} is explicitly given by
\be
    S(R_\AS) \overset{\Huge\mathbf{?}}{\leq} S_\text{bound} = \frac{2\pi r^2_h}{\GN} + \frac{c}{6}\ln\frac{4f(q)}{\kappa^2_h \eps^2}.
    \label{eq:IPforfinitedown}
\ee
We want to check whether this inequality holds.

%-------------------------------------%

\subsubsection*{Without islands}

If we do not take into account the islands for $R_\AS$, the entanglement entropy for~$R_\AS$ has the largest possible value just before the Cauchy surface breaking. For large enough $q$, the entropy exceeds the strong bound (see dark magenta curve in Fig.~\ref{fig:StrongBoundAS}). However, there is some critical value $q_\text{crit}$, such that for $q < q_\text{crit}$ the Cauchy surface breaks before the strong bound is violated, and hence, the information paradox does not arise.

%--------------------------------%

\subsubsection*{Island dominating for finite time}

\begin{figure}[h]\centering
    \includegraphics[width=0.45\textwidth]{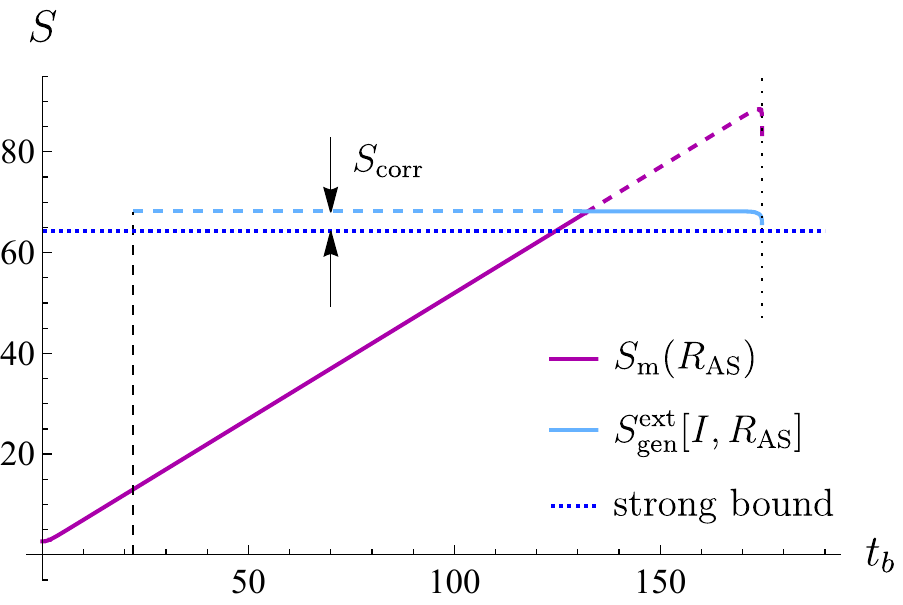}
	\caption{Time evolution of the entropy $S_\m(R_\AS)$ (dark magenta), the generalized entropy $S\extgen[I, R_\AS]$ (sky blue) and the strong bound~\eqref{eq:IPforfinitedown} (dotted blue) for the finite size entangling region $R_\MS$ with $b = 5$ and $q = 350$. The parameters are $r_h = 1$, $c = 3$, $\GN = 0.1$, $\eps = 1$. Introduction of the island helps to keep the entropy close to the bound within the level of corrections $S_\text{corr}$.}
	\label{fig:StrongBoundAS}
\end{figure}

The island contribution to the generalized entropy $S(R_\AS)$ starts to dominate for large enough $q$. It exceeds the strong bound~\eqref{eq:IPFinite} (see Fig.~\ref{fig:StrongBoundAS}), but the soft bound is still satisfied
\be
    S(R_\AS) \leq S_\text{bound} + S_\text{corr},
\ee
where~$S_\text{corr} = S\extgen[I, R_\AS] - S_\text{bound}$ is given by the same expression as in the MS case~\eqref{eq:Scorr_MS} and is proportional to ${c \ll r_h^2/\GN}$.

%-------------------------------------%

\subsubsection*{Never dominant island}

If the island configuration never dominates, then the entropy of matter is always less than the generalized entropy
\be
    S(R_\AS) = S_\m(R_\AS) < S\extgen[I, R_\AS] = S_\text{bound} + S_\text{corr}.
\ee
Therefore, the soft bound~\eqref{eq:IPFinite_soft} is obeyed.
%%%%%%%%%%%%%%%%%%%%%%%%%%%%%%%%%%%%%%%%%%%%%%%%

\section{\label{sec:ConclusionPerspective}Conclusions and future prospects}

In this paper, we address the properties of the entanglement entropy of Hawking radiation on the background of the eternal Schwarzschild black hole in the context of entangling regions of finite size. This can be seen as a starting point for studying the entropy in spacetimes with finite observable domains. Among them are de Sitter~(dS) and Schwarzschild-de Sitter~(SdS) universes, where a physical observer is bounded within the cosmological horizon. Therefore, only finite entangling regions are of physical significance in these spacetimes, and a generalization of the results of this paper on dS and SdS cases is needed.

We have elaborated on the infrared regularization of spacelike infinities of Cauchy surfaces and semi-infinite regions in two-sided Schwarzschild black hole. Our procedure allows to preserve complementarity and pure state condition within the approximation~\eqref{eq:S_N_ints} and in the prescription of subtracting the anomalous term~\eqref{eq:IR_anomaly}.

We have established two qualitatively different types of finite entangling regions --- mirror-symmetric (MS) and asymmetric (AS). The first type represents finite domains between concentric spheres. The slope of the early-time evolution of the entanglement entropy of conformal matter for such regions is twice as large as for the canonical semi-infinite setup. For late times, the entropy saturates at a constant value even without entanglement islands. Island configurations exist only for finite time for such regions.

The outer endpoints of AS regions are chosen spatially symmetric and lie on the same timeslice. The corresponding Cauchy surfaces have a finite lifetime depending on the size of the entangling regions. In this case, the entropy grows linearly. At the time when the Cauchy surface breaks, the entanglement entropy hits singularity. Just before this moment, the island configuration becomes non-symmetric. Also, islands for the regions of both types might never dominate in the generalized entropy functional if their size is relatively small.

We have derived constraints from above on the entanglement entropy for finite entangling regions coming from strong subadditivity and complementarity. The upper bounds depend on the size of the region and on the location of its endpoints. For the AS type, the paradox is solved by entanglement islands. However, the entanglement entropy for MS regions might violate the upper bound even in the island prescription.

%--------------------------------------%

\begin{acknowledgments}
Work of DA, IA and TR, which consisted in designing the project,  calculating entanglement entropy and study of infrared anomalies in the context of entanglement islands (Chapters 1-3 of this paper)  is  supported by the Russian Science Foundation (project 20–12–00200, Steklov Mathematical Institute).  DA, AB and VP are supported by the Foundation for the Advancement of Theoretical Physics and Mathematics “BASIS” (chapters 4-5). The work  was performed at Steklov International Mathematical Center and supported (chapters 4-5) by the Ministry of Science and Higher Education of the Russian Federation (Agreement No. 075-15-2019-1614).
\end{acknowledgments}

%%%%%%%%%%%%%%%%%%%%%%%%%%%%%%%%%%%%%%%%%%%%%%%%

\appendix
\section{\label{sec:appendix}Explicit formulas for entanglement entropy for finite entangling regions}

%----------------------------%

\subsection{MS entangling region without island}

With use of the formula for the entropy collected in multiple intervals~\eqref{eq:S_N_ints} and the formula for distances~\eqref{eq:geod_dist}, along with the definition of up- and down- points~\eqref{eq:up-down}, the entanglement entropy for the mirror-symmetric finite entangling region takes the form
\be
    \begin{aligned}
        S(R_\MS) & = \frac{c}{3}\ln\left(\frac{d(\bb_+^\up,\,\bb_-^\up) \, d(\bq_-^\up,\,\bb_-^\up)}{\eps^2}\right) + \\
        & + \frac{c}{3}\ln\left(\frac{d(\bb_+^\up,\,\bq_+^\up) \, d(\bq_-^\up,\,\bq_+^\up)}{\eps^2}\right) - \\
        & - \frac{c}{3}\ln\left(\frac{d(\bb_-^\up,\,\bq_+^\up) \, d(\bq_-^\up,\,\bb_+^\up)}{\eps^2}\right) = \\
        & = \frac{c}{6}\ln\left[\frac{16f(b)f(q)}{\kappa_h^4\eps^4}\cosh^4 \kappa_h t_b\right] + \\
        & + \frac{c}{3}\ln\left[\frac{\cosh\kappa_h(r_*(q) - r_*(b)) - 1}{\cosh\kappa_h(r_*(q) - r_*(b)) + \cosh 2\kappa_h t_b}\right].
    \end{aligned}
    \label{eq:app:MSnoISl}
\ee

At late times, such that
\be
    \cosh 2\kappa_h t_b \gg \cosh\kappa_h(r_*(q) - r_*(b)),
\ee
the expression simplifies to
\be
    \begin{aligned}
        S(R_\MS)\LateTimes & \simeq \frac{c}{6}\ln\left(\frac{16f(b)f(q)}{\kappa_h^4\eps^4}\right) + \\ 
        & + \frac{c}{3}\ln\Big[\cosh\kappa_h(r_*(q) - r_*(b)) - 1\Big].
    \end{aligned}
    \label{eq:app:MSnoISlLateTime}
\ee

In the limit~$q \to \infty$, we get
\be
    \lim\limits_{q \to \infty} S(R_\MS) = \frac{c}{6}\ln\left(\frac{16f(b)}{\kappa_h^4\eps^4}\cosh^4(\kappa_h t_b)\right).
\ee

%----------------------------%

\subsection{MS entangling region with island}

The entropy for a non-trivial island configuration before the extremization procedure has the form~\eqref{eq:gen_functional}. A general single-interval, non-symmetric ansatz for the island is ${I = [\bp_-,\,\ba_+]}$. The area of its boundary is
\be
    \operatorname{Area}(\partial I) = 4\pi\left(a^2 + p^2\right).
    \label{eq:app:areaterm}
\ee
With use of this, the entanglement entropy functional takes the form
\be
    \begin{aligned}
        & S_\gen[I, R_\MS] = \frac{\pi (a^2 + p^2)}{\GN} + \\
        & + \frac{c}{6}\ln\left[\frac{32\sqrt{f(a)f(p)}f(b)f(q)}{\kappa_h^6\eps^6}\cosh^4\kappa_h t_b\right] + \\
        & + \frac{c}{6}\ln\big[\cosh\kappa_h(r_*(a) - r_*(p)) + \cosh\kappa_h(t_a + t_p)\big] + \\
        & + \frac{c}{6}\ln\left[\frac{\cosh\kappa_h(r_*(b) - r_*(a)) - \cosh\kappa_h(t_a - t_b)}{\cosh\kappa_h(r_*(b) - r_*(a)) + \cosh\kappa_h(t_a + t_b)}\right] + \\
        & + \frac{c}{6}\ln\left[\frac{\cosh\kappa_h(r_*(b) - r_*(p)) - \cosh\kappa_h(t_p - t_b)}{\cosh\kappa_h(r_*(b) - r_*(p)) + \cosh\kappa_h(t_p + t_b)}\right] + \\
        & + \frac{c}{6}\ln\left[\frac{\cosh\kappa_h(r_*(q) - r_*(a)) + \cosh\kappa_h(t_a + t_b)}{\cosh\kappa_h(r_*(q) - r_*(a)) - \cosh\kappa_h(t_a - t_b)}\right] + \\
        & + \frac{c}{6}\ln\left[\frac{\cosh\kappa_h(r_*(q) - r_*(p)) + \cosh\kappa_h(t_p + t_b)}{\cosh\kappa_h (r_*(q) - r_*(p)) - \cosh\kappa_h(t_p - t_b)}\right] + \\
        & + \frac{c}{3}\ln\left[\frac{\cosh\kappa_h(r_*(q) - r_*(b)) - 1}{\cosh\kappa_h(r_*(q) - r_*(b)) + \cosh 2\kappa_h t_b}\right].
    \end{aligned}
    \label{eq:app:MSwISl}
\ee
This expression is to be extremized with respect to the parameters $(a, t_a, p, t_p)$.

The four-parameter extremization can be brought to extremization over only two parameters. Indeed, all the terms in $S_\gen[I, R_\MS]$ are symmetric with respect to permutations ${a \leftrightarrow p}$, ${t_a \leftrightarrow t_p}$, hence, it is reasonable to take a mirror-symmetric ansatz for the island, ${I_\MS = [\ba_-,\,\ba_+]}$. For such a choice, the full expression for the generalized entropy functional reads as
\be
    \begin{aligned}
        & S_\gen[I_\MS, R_\MS] = \frac{2\pi a^2}{\GN} + \\
        & + \frac{c}{6}\ln\left[\frac{64 f(a)f(b) f(q)}{\kappa^6_h\eps^6}\cosh^2\kappa_h t_a \cosh^4\kappa_h t_b\right] + \\
        & + \frac{c}{3}\ln\left[\frac{\cosh\kappa_h(r_*(b) - r_*(a)) - \cosh\kappa_h(t_a - t_b)}{\cosh\kappa_h(r_*(b) - r_*(a)) + \cosh\kappa_h(t_a + t_b)}\right] + \\
        & + \frac{c}{3}\ln\left[\frac{\cosh\kappa_h(r_*(q) - r_*(a)) + \cosh\kappa_h(t_a + t_b)}{\cosh\kappa_h(r_*(q) - r_*(a)) - \cosh\kappa_h(t_a - t_b)}\right] + \\
        & + \frac{c}{3}\ln\left[\frac{\cosh\kappa_h(r_*(q) - r_*(b)) - 1}{\cosh\kappa_h(r_*(q) - r_*(b)) + \cosh 2\kappa_h t_b}\right].
    \end{aligned}
    \label{eq:app:qupfiniteentropy2}
\ee
In this expression, we can identify a part of the terms with the generalized entropy functional for the semi-infinite region and the other part with an effect of finite size of the region,
\be
    \begin{aligned}
        & S_\gen[I_\MS, R_\MS] = S_\gen[I_\MS, R_\infty] + \\
        & + \frac{c}{6}\ln\left[\frac{4f(q)}{\kappa^2_h\eps^2}\cosh^2\kappa_h t_b\right] + \\
        & + \frac{c}{3}\ln\left[\frac{\cosh\kappa_h(r_*(q) - r_*(a)) + \cosh\kappa_h(t_a + t_b)}{\cosh\kappa_h(r_*(q) - r_*(a)) - \cosh\kappa_h(t_a - t_b)}\right] + \\
        & + \frac{c}{3}\ln\left[\frac{\cosh\kappa_h(r_*(q) - r_*(b)) - 1}{\cosh\kappa_h(r_*(q) - r_*(b)) + \cosh 2\kappa_h t_b}\right].
    \end{aligned}
\ee

It can be simplified further by using small-, intermediate- or large-time approximations.

We use a general property that for ${x \gg 1}$ holds ${\cosh x \simeq \nicefrac{1}{2}\,e^x}$, and for $A$ and $B$ such that ${\cosh B \ll \cosh A}$ and $A \gg 1$ holds
\be
    \begin{aligned}
        & \ln(\cosh A + \cosh B) \approx A + 2\cosh B\exp(-A).
    \end{aligned}
    \label{eq:app:simplify}
\ee

Then, for early times (small with respect to tortoise radial coordinate of the inner endpoints),
\be
    \begin{aligned}
        \cosh\kappa_h(t_a \pm t_b) & \ll \cosh\kappa_h(r_*(b) - r_*(a)) < \\
        & < \cosh\kappa_h(r_*(q) - r_*(a)),
    \end{aligned}
\ee
we obtain
\be
    \begin{aligned}
        S_\gen[I_\MS, R_\MS]\EarlyTimes & \simeq S_\gen[I_\MS, R_\infty] + \\
        & + \frac{c}{6}\ln\left(\frac{4f(q)}{\kappa^2_h\eps^2}\cosh^2\kappa_h t_b\right).
    \end{aligned}
    \label{eq:app:MSwIslEarlyTime}
\ee
Since the remaining term, describing the effect of the finite size of the region, does not depend neither on~$a$ nor~$t_a$, there is no solution as in the semi-infinite case~\cite{HIM}.

For late times
\be
    \begin{aligned}
        \cosh\kappa_h(t_a + t_b) & \gg \cosh\kappa_h(r_*(q) - r_*(a)), \\
        \cosh 2\kappa_h t_b & \gg \cosh\kappa_h(r_*(q) - r_*(b)),
    \end{aligned}
\ee
the functional reduces to
\be
    \begin{aligned}
	    & S_\gen[I_\MS, R_\MS]\LateTimes \simeq \frac{2\pi a^2}{\GN} + \\
	    & + \frac{c}{3}\,\kappa_h(t_a + t_b) + \frac{c}{6}\ln\left[\frac{4 f(a)f(b)f(q)}{\kappa^6_h\eps^6}\cosh^2\kappa_h t_b\right] + \\
	    & + \frac{c}{3}\ln\left[\frac{\cosh\kappa_h(r_*(b) - r_*(a)) - \cosh\kappa_h(t_a - t_b)}{\cosh\kappa_h(r_*(q) - r_*(a)) - \cosh\kappa_h(t_a - t_b)}\right] + \\
	    & + \frac{c}{3}\ln\Big[\cosh\kappa_h(r_*(q) - r_*(b)) - 1\Big].
    \end{aligned}
    \label{eq:app:MSwIslLargeTime}
\ee
Due to linear growth, there is no solution to the extremization equation for $t_a \approx t_b$.

For intermediate times
\be
    \begin{aligned}
        & 1 \ll \cosh\kappa_h (r_*(b) - r_*(a)) \ll x \ll \\
        & \ll \cosh\kappa_h (r_*(q) - r_*(b)) < \cosh\kappa_h (r_*(q) - r_*(a)),
    \end{aligned}
    \label{eq:app:conditiononislandUP}
\ee
where $x = \cosh\kappa_h(t_a + t_b) \text{ or } \cosh 2\kappa_ht_b$, making use of~\eqref{eq:app:simplify} and $t_a,\,t_b \gg r_h$ we can simplify the generalized entropy functional to
\be
    \begin{aligned}
        & S_\gen[I_\MS, R_\MS]\InterTimes \simeq \frac{2\pi a^2}{\GN} + \\
        & + \frac{c}{6}\ln\left(\frac{4f(a)f(b)f(q)}{\kappa^6_h\eps^6}\cosh^2\kappa_h t_b\right) + \\
        & + \frac{c}{3}\kappa_h(r_*(b) - r_*(a)) - \\
        & - \frac{2c}{3}\cosh\kappa_h(t_a - t_b)e^{- \kappa_h(r_*(b) - r_*(a))} - \\
        & - \frac{c}{3}e^{-\kappa_h (t_a + t_b) + \kappa_h(r_*(b) - r_*(a))} - \\
        & - \frac{c}{6}e^{2\kappa_h t_b - \kappa_h(r_*(q) - r_*(b))} + \\
        & + \frac{c}{3}\cosh\kappa_h(t_a - t_b)e^{- \kappa_h(r_*(q) - r_*(a))} + \\
        & + \frac{c}{6}e^{\kappa_h (t_a + t_b) - \kappa_h(r_*(q) - r_*(a))}.
    \end{aligned}
\ee
The extremal curve with respect to time coordinate is at $t_a = t_b$. Using this, we get
\be
    \begin{aligned}
        & S_\gen[I_\MS, R_\MS]\InterTimes \simeq \frac{2\pi a^2}{\GN} + \\
        & + \frac{c}{6}\ln\left(\frac{4f(a)f(b)f(q)}{\kappa^6_h\eps^6}\cosh^2\kappa_h t_b\right) + \\
        & + \frac{c}{3}\kappa_h(r_*(b) - r_*(a)) - \frac{c}{3}e^{2\kappa_h t_b - \kappa_h(r_*(q) - r_*(b))}.
    \end{aligned}
\ee
We keep the vanishingly small last term from other exponential terms as that describing the leading order effect of the finite size of the region.

Taking then near-horizon-zone ansatz for the radial coordinate of the island, ${a = r_h + \delta a}$, ${\delta a \ll r_h}$, we find the analytical expression~\eqref{eq:RmsAnFormEn} for the entropy in the leading order in $\delta a/r_h$ without solving extremization equation on $a$ explicitly,
\be
    \begin{aligned}
        & S\extgen[I_\MS, R_\MS]\InterTimes \simeq \\
        & \simeq \frac{2\pi r_h^2}{\GN} + \frac{c}{6}\ln\frac{f(b)}{\kappa^4_h\eps^4} + \frac{c}{3}\kappa_h r_*(b) - \frac{c}{6} + \\
        & + \frac{c}{6}\ln\left(\frac{4f(q)\cosh^2\kappa_h t_b}{\kappa^2_h\eps^2}\right) - \frac{c}{3}e^{2\kappa_h t_b - \kappa_h(r_*(q) - r_*(b))} + \\
        & + O(\delta a/r_h).
    \end{aligned}
    \label{eq:app:RmsAnFormEn}
\ee

%----------------------------%

\subsection{AS entangling region without island}

The entanglement entropy for the asymmetric finite entangling region takes the form
\be
    \begin{aligned}
        & S(R_\AS) = \frac{c}{3}\ln\left(\frac{d(\bb_+^\up,\,\bb_-^\up) \, d(\bq_-^\down,\,\bb_-^\up)}{\eps^2}\right) + \\
        & + \frac{c}{3}\ln\left(\frac{d(\bb_+^\up,\,\bq_+^\up) \, d(\bq_-^\down,\,\bq_+^\up)}{\eps^2}\right) - \\
        & - \frac{c}{3}\ln\left(\frac{d(\bb_-^\up,\,\bq_+^\up) \, d(\bq_-^\down,\,\bb_+^\up)}{\eps^2}\right) = \\
        & = \frac{c}{6}\ln\left[\frac{16f(b)f(q)}{\kappa^4_h\eps^4}\cosh^2\kappa_h t_b\right] + \\
        & + \frac{c}{6}\ln\left[\frac{\cosh\kappa_h(r_*(q) - r_*(b)) - \cosh 2 \kappa_h t_b}{\cosh\kappa_h(r_*(q) - r_*(b)) + \cosh 2\kappa_h t_b}\right] + \\
        & + \frac{c}{6}\ln\left[\frac{\cosh\kappa_h(r_*(q) - r_*(b)) - 1}{\cosh\kappa_h(r_*(q) - r_*(b)) + 1}\right].
    \end{aligned}
    \label{eq:app:ASnoISl}
\ee

In the limit $q \to \infty$, we obtain
\be
    \begin{aligned}
        & \lim\limits_{q \to \infty} S(R_\AS) = \frac{c}{6}\ln\left(\frac{16f(b)}{\kappa^4_h\eps^4}\cosh^2\kappa_h t_b\right).
    \end{aligned}
\ee

%----------------------------%

\subsection{AS entangling region with island}

The generalized entropy functional for a general non-symmetric island ansatz is
\be
    \begin{aligned}
        & S_\gen[I, R_\AS] = \frac{\pi (a^2 + p^2)}{\GN} + \\
        & + \frac{c}{6}\ln\left[\frac{32\sqrt{f(a)f(p)}f(b)f(q)}{\kappa_h^6\eps^6}\cosh^2\kappa_h t_b\right] + \\
        & + \frac{c}{6}\ln\big[\cosh\kappa_h(r_*(a) - r_*(p)) + \cosh\kappa_h(t_a + t_p)\big] + \\
        & + \frac{c}{6}\ln\left[\frac{\cosh\kappa_h(r_*(b) - r_*(a)) - \cosh\kappa_h(t_a - t_b)}{\cosh\kappa_h(r_*(b) - r_*(a)) + \cosh\kappa_h(t_a + t_b)}\right] + \\
        & + \frac{c}{6}\ln\left[\frac{\cosh\kappa_h(r_*(b) - r_*(p)) - \cosh\kappa_h(t_p - t_b)}{\cosh\kappa_h(r_*(b) - r_*(p)) + \cosh\kappa_h(t_p + t_b)}\right] + \\
        & + \frac{c}{6}\ln\left[\frac{\cosh\kappa_h(r_*(q) - r_*(b)) - 1}{\cosh\kappa_h(r_*(q) - r_*(b)) + 1}\right] + \\
        & + \frac{c}{6}\ln\left[\frac{\cosh\kappa_h(r_*(q) - r_*(b)) - \cosh 2\kappa_h t_b}{\cosh\kappa_h(r_*(q) - r_*(b)) + \cosh 2\kappa_h t_b}\right] + \\
        & + \frac{c}{6}\ln\left[\frac{\cosh\kappa_h(r_*(q) - r_*(a)) + \cosh\kappa_h(t_a - t_b)}{\cosh\kappa_h(r_*(q) - r_*(a)) - \cosh\kappa_h(t_a - t_b)}\right] + \\
        & + \frac{c}{6}\ln\left[\frac{\cosh\kappa_h(r_*(q) - r_*(p)) + \cosh\kappa_h(t_p + t_b)}{\cosh\kappa_h(r_*(q) - r_*(p)) - \cosh\kappa_h(t_p + t_b)}\right].
    \end{aligned}
    \label{eq:app:ASwISl}
\ee

The only non-symmetric under exchange ${a \leftrightarrow p}$, ${t_a \leftrightarrow t_p}$ terms in this expression are in the last two lines. These terms are vanishingly small for times such that
\be
    \begin{aligned}
        & \cosh\kappa_h(t_p + t_b) \ll \cosh\kappa_h(r_*(q) - r_*(p)), \\
        & \cosh\kappa_h(t_a - t_b) \ll \cosh\kappa_h(r_*(q) - r_*(a)).
    \end{aligned}
\ee
Under these assumptions taking the mirror-symmetric ansatz for island ${I_\MS = [\ba_-,\,\ba_+]}$ is valid. Substituting it to~$S[I, R_\AS]$, we arrive at
\be
    \begin{aligned}
        & S_\gen[I_\MS, R_\AS] \simeq \frac{2\pi a^2}{\GN} + \\
        & + \frac{c}{6}\ln\left[\frac{64f(a)f(b)f(q)}{\kappa^6_h\eps^6}\cosh^2\kappa_h t_a\cosh^2\kappa_h t_b\right] + \\
        & + \frac{c}{3}\ln\left[\frac{\cosh\kappa_h(r_*(b) - r_*(a)) - \cosh\kappa_h(t_a - t_b)}{\cosh\kappa_h(r_*(b) - r_*(a)) + \cosh\kappa_h(t_a + t_b)}\right] + \\
        & + \frac{c}{6}\ln\left[\frac{\cosh\kappa_h(r_*(q) - r_*(b)) - 1}{\cosh\kappa_h(r_*(q) - r_*(b)) + 1}\right] + \\
        & + \frac{c}{6}\ln\left[\frac{\cosh\kappa_h(r_*(q) - r_*(b)) - \cosh 2\kappa_h t_b}{\cosh\kappa_h(r_*(q) - r_*(b)) + \cosh 2\kappa_h t_b}\right] + \\
        & + \frac{c}{6}\ln\left[\frac{\cosh\kappa_h(r_*(q) - r_*(a)) + \cosh\kappa_h(t_a - t_b)}{\cosh\kappa_h(r_*(q) - r_*(a)) - \cosh\kappa_h(t_a - t_b)}\right] + \\
        & + \frac{c}{6}\ln\left[\frac{\cosh\kappa_h(r_*(q) - r_*(a)) + \cosh\kappa_h(t_a + t_b)}{\cosh\kappa_h(r_*(q) - r_*(a)) - \cosh\kappa_h(t_a + t_b)}\right].
    \end{aligned}
    \label{eq:app:qdownfiniteentropy3}
\ee

If we then assume that the island endpoints are near the horizon and times are long before the Cauchy surface breaking~\eqref{eq:breaktime} (intermediate times), such that
\be
    \begin{aligned}
        & 1 \ll \cosh\kappa_h(r_*(b) - r_*(a))  \ll x \ll \\
        & \ll \cosh\kappa_h(r_*(q) - r_*(b)) < \cosh\kappa_h(r_*(q) - r_*(a)),
    \end{aligned}
    \label{eq:app:uslAS}
\ee
where $x = \cosh\kappa_h(t_a + t_b) \text{ or } \cosh 2\kappa_ht_b$, making use of~\eqref{eq:app:simplify} and $t_a,\,t_b \gg r_h$ we can simplify the generalized entropy functional to
\be
    \begin{aligned}
        & S_\gen[I_\MS, R_\AS]\InterTimes \simeq \frac{2\pi a^2}{\GN} + \\
        & + \frac{c}{6}\ln\left(\frac{4f(a)f(b)f(q)}{\kappa^6_h\eps^6}\right) + \frac{c}{3}\kappa_h(r_*(b) - r_*(a)) - \\
        & - \frac{2c}{3}\cosh\kappa_h(t_a - t_b)e^{- \kappa_h(r_*(b) - r_*(a))} - \\
        & - \frac{c}{3}e^{-\kappa_h (t_a + t_b) + \kappa_h(r_*(b) - r_*(a))} - \\
        & - \frac{c}{3}e^{2\kappa_h t_b - \kappa_h(r_*(q) - r_*(b))} + \\
        & + \frac{2c}{3}\cosh\kappa_h(t_a - t_b)e^{- \kappa_h(r_*(q) - r_*(a))} + \\
        & + \frac{c}{3}e^{\kappa_h (t_a + t_b) - \kappa_h(r_*(q) - r_*(a))}.
    \end{aligned}
\ee
Extremizing with respect to $t_a$, we obtain
\be
    \begin{aligned}
        & S_\gen[I_\MS, R_\AS]\InterTimes \simeq \frac{2\pi a^2}{\GN} + \frac{c}{6}\ln\left(\frac{4f(a)f(b)f(q)}{\kappa^6_h\eps^6}\right) + \\
        & + \frac{c}{3}\kappa_h(r_*(b) - r_*(a)) - \frac{c}{3}e^{2\kappa_h t_b - \kappa_h(r_*(q) - r_*(b))},
    \end{aligned}
\ee
and hence an approximate analytical expression~\eqref{eq:RasAnFormEn} in the near-horizon zone in the leading order in $\delta a/r_h$ is
\be
    \begin{aligned}
        & S\extgen[I_\MS, R_\AS]\InterTimes \simeq \\
        & \simeq \frac{2\pi r_h^2}{\GN} + \frac{c}{6}\ln\frac{f(b)}{\kappa^4_h\eps^4} + \frac{c}{3}\kappa_h r_*(b) - \frac{c}{6} + \\
        & + \frac{c}{6}\ln\frac{4f(q)}{\kappa^2_h\eps^2} - \frac{c}{3}e^{2\kappa_h t_b - \kappa_h(r_*(q) - r_*(b))} + O(\delta a/r_h).
    \end{aligned}
    \label{eq:app:ASwISlInterTime}
\ee

%----------------------------%

\subsection{Outer entangling region for AS finite region}

The entanglement entropy for $C_\AS$ takes the form
\be
    \begin{aligned}
        S_\m(C_\AS) & = \lim\limits_{w \to \infty}\left[\frac{c}{3}\ln\left(\frac{d(\bq_+^\up,\,\bq_-^\down) \, d(\bq_+^\up,\,\bw_+^\up)}{\eps^2}\right)\right. + \\
        & + \frac{c}{3}\ln\left(\frac{d(\bw_-^\down,\,\bq_-^\down) \, d(\bw_-^\down,\,\bw_+^\up)}{\eps^2}\right) - \\
        & - \left.\frac{c}{3}\ln\left(\frac{d(\bw_-^\down,\,\bq_+^\up) \, d(\bw_+^\up,\,\bq_-^\down)}{\eps^2}\right)\right] = \\
        & = \lim\limits_{w \to \infty}\frac{c}{6}\ln\left(\frac{16f(q)f(w)}{\kappa_h^4\eps^4}\right) + \\
        & + \lim\limits_{w \to \infty}\frac{c}{3}\ln\left(\frac{\cosh\kappa_h(r_*(w) - r_*(q)) - 1}{\cosh\kappa_h(r_*(w) - r_*(q)) + 1}\right) = \\
        & = \frac{c}{6}\ln\frac{4f(q)}{\kappa_h^2\eps^2} + \frac{c}{3}\ln\frac{2}{\kappa_h\eps}.
    \end{aligned}
    \label{eq:app:CAS}
\ee
According to our prescription for IR regularization, the latter term is to be subtracted. Interestingly, the same result can be obtained if $C_\AS$ is considered as the complement of the regularized Cauchy surface. The entropy for it we have already calculated in~\eqref{eq:reg-Cauchy-gen}, $S_\m(C_\AS) = S_\m\left(\overline{\Sigma}_\text{reg}\right)$.

%%%%%%%%%%%%%%%%%%%%%%%%%%%%%%%%%%%%%%%%%%%%%%%%

\bibliography{IRregSchwBH}

\end{document}